\documentclass[12pt]{article}
\usepackage{a4wide,epsfig,amsmath,amssymb,cite,scalefnt}

\newcommand{\drbar}{$\overline{\mbox{\tt DR}}$}
\newcommand{\msbar}{$\overline{\mbox{\tt MS}}$}
\newcommand{\drbarf}{\overline{\rm DR}}
\newcommand{\msbarf}{\overline{\rm MS}}
\newcommand{\asDRbar}{\alpha_s^{\overline{\rm DR}}}
\newcommand{\asMSbar}{\alpha_s^{\overline{\rm MS}}}

\newcommand{\apiDR}{\frac{\asDRbar}{\pi}}

\newcommand{\aepi}{\frac{\alpha_e}{\pi}}

\newcommand{\mMSbar}{m^{\overline{\rm MS}}}

\newcommand{\dreg}{{\tt DREG}}
\newcommand{\dred}{{\tt DRED}}

\newcommand{\qsla}{q\!\!\!/\,\,}
\newcommand{\Qsla}{Q\!\!\!\!/\,\,\,}
\newcommand{\Lmu}{L_\mu}
\newcommand{\ep}{\epsilon}
\newcommand{\order}[1]{\mathcal{O}\left( #1 \right)}

\begin{document}

\title{\vskip-3cm{\baselineskip14pt
    \begin{flushleft}
      \normalsize SFB/CPP-07-04 \\
      \normalsize TTP07-04  \\
      \normalsize DESY07-016  \\
      \normalsize hep-ph/0702185
  \end{flushleft}}
  \vskip1.5cm
  Relation between the pole and the minimally subtracted mass 
  in dimensional regularization and dimensional reduction
  to three-loop order
}
\author{\small P. Marquard$^{(a)}$, L. Mihaila$^{(a)}$, 
  J.H. Piclum$^{(a,b)}$ and M. Steinhauser$^{(a)}$\\[1em]
{\small\it (a) Institut f{\"u}r Theoretische Teilchenphysik,
  Universit{\"a}t Karlsruhe (TH)}\\
{\small\it 76128 Karlsruhe, Germany}
\\
{\small\it (b) II. Institut f\"ur Theoretische Physik, 
  Universit\"at Hamburg}\\
{\small\it 22761 Hamburg, Germany}
}

\date{}

\maketitle

\thispagestyle{empty}

\begin{abstract}
  We compute the relation between the pole quark mass and the minimally
  subtracted quark mass in the framework of QCD applying dimensional
  reduction as a regularization scheme. Special emphasis is put on the
  evanescent couplings and the renormalization of the
  $\varepsilon$-scalar mass.
  As a by-product we obtain the three-loop on-shell renormalization
  constants $Z_m^{\rm OS}$ and $Z_2^{\rm OS}$ in dimensional
  regularization and thus provide the
  first independent check of the analytical results computed several
  years ago.
\medskip


\end{abstract}

\thispagestyle{empty}

\newpage


\section{\label{sec::intro}Introduction}

In quantum chromodynamics (QCD), like in any other renormalizeable quantum
field theory, it is crucial to specify the precise meaning of the
parameters appearing in the underlying Lagrangian --- in particular
when higher order quantum corrections are considered.
The canonical choice for the coupling constant of QCD, $\alpha_s$, is
the so-called modified minimal subtraction ($\overline{\rm MS}$)
scheme~\cite{Bardeen:1978yd} which has the advantage that the beta function,
ruling the scale dependence of the coupling, is mass-independent.
On the other hand, for a heavy quark besides the
$\overline{\rm MS}$ scheme also other definitions are important ---
first and foremost the pole mass. Whereas the former definition is
appropriate for those processes where the relevant energy scales are
much larger than the quark mass the pole mass is the relevant
definition for threshold processes.
Thus it is important to have precise conversion formulae at hand in order
to convert one definition into the other.

By far the most loop calculations performed within QCD are based on 
dimensional regularization (\dreg{}) in order to handle the infinities
which occur in intermediate steps. It is well known, however, that 
it is not convenient to apply \dreg{} to supersymmetric theories since
it introduces a mismatch  between the numbers of fermionic and
bosonic degrees of freedom in super-multiplets. 
In order to circumvent this problem and, at the same time,
take over as many advantages
as possible from \dreg{} the regularization scheme dimensional
reduction (\dred{}) has been invented (see, e.g.,
Refs.~\cite{Siegel:1979wr,Capper:1979ns}). 
Indeed, the application of \dred{} to supersymmetric theories leads to
a relatively small price one has to pay at the technical level.
An elegant way is to introduce an additional scalar particle 
(the so-called $\varepsilon$ scalar) at the
level of the Lagrangian and to proceed for the practical calculation
of the Feynman diagrams as in \dreg{}. The situation becomes more
complicated in case the symmetry between fermions and bosons in the
underlying theory is distorted, e.g., after some heavy squarks have
been integrated out. In such situations couplings involving
the $\varepsilon$ scalar, which are called evanescent couplings, renormalize
differently from the gauge couplings and therefore one has to allow for 
new couplings in the theory.
The same is true if \dred{} is applied to QCD: in addition to $\alpha_s$
four new  
couplings have to be introduced, each of which has their own beta
function governing both the running and the renormalization. In this
paper we take over the notation
from~\cite{Harlander:2006xq} and denote them by $\alpha_e$ (the coupling
between $\varepsilon$ scalars and quarks) and $\eta_i$ ($i=1,2,3$), which
describe three different four-$\varepsilon$ vertices.

The one-loop relation between the \msbar{} and pole mass has been
considered long ago in Ref.~\cite{Tarrach:1980up}. At the beginning of
the nineties the mass relation to two-loop order~\cite{Gray:1990yh} 
was one of the first applications of two-loop on-shell
integrals. In a subsequent paper also the two-loop result for the
on-shell wave function counterterm has been obtained~\cite{Broadhurst:1991fy}.
The three-loop mass relation has been computed for the first
time in Ref.~\cite{Chetyrkin:1999ys,Chetyrkin:1999qi} where
the off-shell fermion propagator has been considered for small and
large external momenta. The on-shell quantities have been 
obtained with the help of a conformal mapping and Pad\'e
approximation. The numerical results of
Ref.~\cite{Chetyrkin:1999ys,Chetyrkin:1999qi} have later 
been confirmed in Ref.~\cite{Melnikov:2000qh} by an analytical on-shell
calculation. The three-loop result for $Z_2^{\rm OS}$ has been
obtained in Ref.~\cite{Melnikov:2000zc}.
In this paper we provide the first independent check of the 
analytical results for $Z_m^{\rm OS}$ and $Z_2^{\rm OS}$ in the
\dreg{} scheme.

The renormalization scheme based on \dred{} together with modified minimal
subtraction is called \drbar{}.
As far as the relation between the \drbar{} and the pole mass is
concerned one can find the one-loop result in Ref.~\cite{Martin:1993yx}. 
The two-loop calculation~\cite{Avdeev:1997sz} has been performed in
\dred{}, identifying, 
however, the evanescent coupling $\alpha_e$ with $\alpha_s$.
In this paper we will provide the more general result and furthermore
extend the relation to three-loop order.

The paper is organized as follows: In Section~\ref{sec::dimreg}
we outline the general strategy, provide technical details 
and derive the result for the on-shell mass and wave function
counterterm in the case of
dimensional regularization. The peculiarities of dimensional
reduction are explained in Section~\ref{sec::ren_dred} where all
relevant counterterms are listed. Finally, Section~\ref{sec::dimred}
contains the main result of our paper: the relation between the pole
mass and the minimally subtracted mass in the framework of dimensional
reduction up to three-loop order.


\section{\label{sec::dimreg}
  On-shell counterterms for mass and wave function
  in dimensional regularization}

In order to compute the counterterms
for the quark mass and wave function 
one has to put certain requirements on the 
pole and the residual of the quark propagator.
More precisely, in the on-shell scheme we demand that
the quark two-point function has a zero at the position of the
on-shell mass and that the residual of the propagator is $-i$.
Thus, the renormalized quark propagator is given by
\begin{eqnarray}
  S_F(q) &=& \frac{-i Z_2^{\rm OS}}{\qsla - m_{q,0} + \Sigma(q,M_q)}
  \label{eq::defs::fullprop1} \\
  &\stackrel{q^2\to M_q^2}{\longrightarrow}& \frac{-i}{\qsla - M_q}\,,
  \label{eq::defs::fullprop2}
\end{eqnarray}
where the renormalization constants are defined as
\begin{eqnarray}
  m_{q,0} &=& Z_m^{\rm OS}\, M_q\,,
  \label{eq::defs::defZm} \\
  \psi_0 &=& \sqrt{Z_2^{\rm OS}}\, \psi\,.
  \label{eq::defs::defZ2}
\end{eqnarray}
$\psi$ is the quark field with mass $m_q$, $M_q$ is the 
on-shell mass and bare quantities are denoted by a subscript 0.
$\Sigma$ denotes the quark self-energy
contributions which can be decomposed as
\begin{eqnarray}
  \Sigma(q,m_q) &=&
  m_q\, \Sigma_1(q^2,m_q) + (\qsla - m_q)\, \Sigma_2(q^2,m_q)\,.
  \label{eq::defs::sigmadecomp}
\end{eqnarray}

The calculation outlined in Ref.~\cite{Gray:1990yh} for the evaluation of
$Z_m^{\rm OS}$ and $Z_2^{\rm OS}$ reduces all occurring Feynman diagrams
to the evaluation of on-shell integrals at the bare mass scale. In
particular, it avoids the introduction of explicit counterterm diagrams.
At three-loop order we find it more convenient to follow the more
direct approach described in
Ref.~\cite{Melnikov:2000qh,Melnikov:2000zc}, which requires 
the calculation of diagrams with mass counterterm insertion. Following
the latter reference, we expand $\Sigma$ around $q^2 = M_q^2$
\begin{eqnarray}
  \Sigma(q,M_q) &\approx& M_q\, \Sigma_1(M_q^2,M_q) + (\qsla - M_q)\,
  \Sigma_2(M_q^2,M_q) \nonumber\\
  && + M_q \frac{\partial}{\partial q^2} \Sigma_1(q^2,M_q) \Big|_{q^2 =
  M_q^2} (q^2 - M_q^2) + \dots \nonumber\\
  &\approx& M_q\, \Sigma_1(M_q^2,M_q) \nonumber\\ 
  && + (\qsla - M_q) \left( 2M_q^2 \frac{\partial}{\partial q^2}
  \Sigma_1(q^2,M_q) \Big|_{q^2 = M_q^2} + \Sigma_2(M_q^2,M_q) \right) +
  \dots\,.
  \label{eq::defs::taylor}
\end{eqnarray}
Inserting Eq.~(\ref{eq::defs::taylor}) into
Eq.~(\ref{eq::defs::fullprop1}) and comparing to
Eq.~(\ref{eq::defs::fullprop2}) we find the following formulae for the
renormalization constants
\begin{eqnarray}
  Z_m^{\rm OS} &=& 1 + \Sigma_1(M_q^2,M_q)\,,
  \label{eq::defs::calcZm} \\
  \left( Z_2^{\rm OS} \right)^{-1} &=& 1 + 2M_q^2
  \frac{\partial}{\partial q^2} \Sigma_1(q^2,M_q) \Big|_{q^2 = M_q^2} +
  \Sigma_2(M_q^2,M_q) \,.
  \label{eq::defs::calcZ2}
\end{eqnarray}

If we consider the external momentum of the quarks to be $q = Q(1+t)$
with $Q^2 = M_q^2$, the self-energy can be written as
\begin{equation}
  \Sigma(q,M_q) = M_q \Sigma_1(q^2,M_q) + (\Qsla - M_q)
  \Sigma_2(q^2,M_q) + t\Qsla \Sigma_2(q^2,M_q)\,.
  \label{eq::defs::seldecompt}
\end{equation}
Let us now consider the quantity ${\rm Tr} \{ \frac{\Qsla + M_q}{4M_q^2}
\Sigma \}$ and expand to first order in $t$
\begin{eqnarray}
  {\rm Tr} \left\{ \frac{\Qsla + M_q}{4M_q^2} \Sigma(q,M_q) \right\} &=&
  \Sigma_1(q^2,M_q) + t \Sigma_2(q^2,M_q) \nonumber \\
  &=& \Sigma_1(M_q^2,M_q)
  + \left( 2M_q^2 \frac{\partial}{\partial q^2} \Sigma_1(q^2,M_q)
  \Big|_{q^2 = M_q^2} \!\!+\! \Sigma_2(M_q^2,M_q) \right) t \nonumber\\ 
  &&  + \order{t^2} \,.
  \label{eq::defs::trace}
\end{eqnarray}
Thus, to obtain $Z_m^{\rm OS}$ one only needs to calculate 
$\Sigma_1$ for $q^2 = M_q^2$. 
To calculate $Z_2^{\rm OS}$, one has to compute the first derivative of
the self-energy diagrams. The mass renormalization is taken into account
iteratively by calculating one- and two-loop diagrams with zero-momentum
insertions.

\begin{figure}[t]
  \leavevmode
  \epsfxsize=\textwidth
  \epsffile{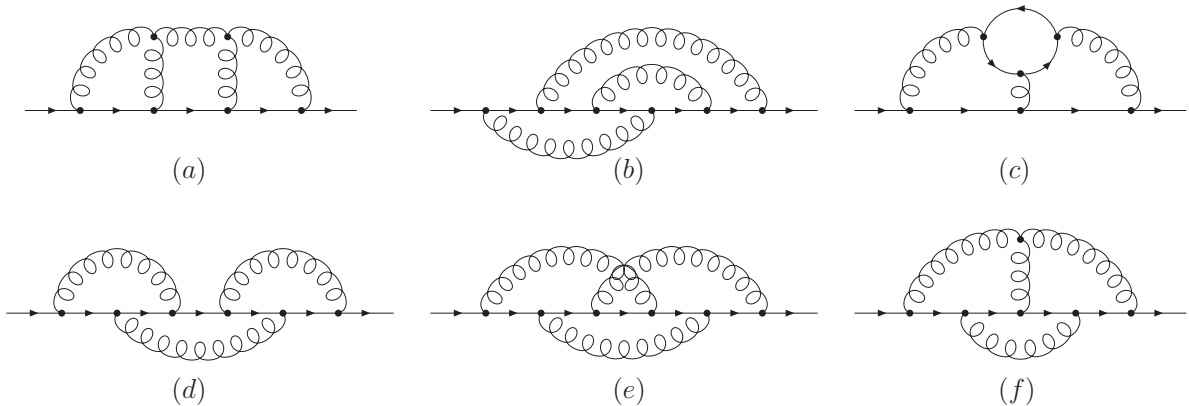}
\caption{\label{fig::selfdias} Sample three-loop diagrams. Solid lines
  denote massive quarks with mass $m_q$ and curly lines denote
  gluons. In
  the closed fermion loops all quark flavours have to be considered.}
\end{figure}

Sample diagrams contributing to the quark propagator
are shown in Fig.~\ref{fig::selfdias}.
All occurring Feynman integrals can be mapped onto $J_+^{(3)}$ and
$L_+^{(3)}$ as given in Eq.~(4) of Ref.~\cite{Marquard:2006qi} 
and to seven more similar ones which will be listed in
Ref.~\cite{Piclum_diss}.

All Feynman diagrams are generated with {\tt
QGRAF}~\cite{Nogueira:1991ex} and the various topologies are identified 
with the help of {\tt q2e} and {\tt
exp}~\cite{Harlander:1997zb,Seidensticker:1999bb}.
In a next step the reduction of the various functions to so-called master
integrals has to be achieved. For this step we use the so-called
Laporta method~\cite{Laporta:1996mq,Laporta:2001dd} 
which reduces the three-loop integrals to 19
master integrals. We use the implementation of Laporta's
algorithm in the program {\tt Crusher}~\cite{PMDS}. 
It is written in {\tt C++} and uses
{\tt GiNaC}~\cite{Bauer:2000cp} for simple
manipulations like taking derivatives of polynomial quantities. In the
practical implementation of the Laporta algorithm one of the most
time-consuming operations is the simplification of the coefficients
appearing in front of the individual integrals. This task is performed
with the help of {\tt Fermat}~\cite{fermat} where a special interface
has been used (see Ref.~\cite{Tentyukov:2006ys}).
The main features of the
implementation are the automated generation of the
integration-by-parts (IBP) identities~\cite{Chetyrkin:1981qh} 
and a complete symmetrization of the diagrams.

\begin{figure}
  \leavevmode
  \epsfxsize=\textwidth
  \epsffile{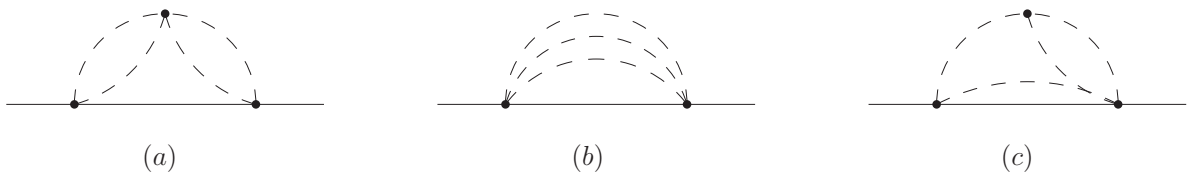}
\caption{\label{fig::selfmaster} Three-loop master integrals. Solid
  lines denote massive and dashed lines are massless scalar propagators.}
\end{figure}

The results for the master integrals can be found in
Ref.~\cite{Melnikov:2000zc}. We have checked the results numerically
with the Mellin-Barnes method~\cite{Smirnov:1999gc,Tausk:1999vh} and the
program {\tt MB}~\cite{Czakon:2005rk}. We did, however, find some
differences to 
Ref.~\cite{Melnikov:2000zc}: In addition to the 18 master integrals
given in that reference, we found the integral depicted in
Fig.~\ref{fig::selfmaster}$(a)$ where the result can be found in
Eq.~(A.2) of Ref.~\cite{Marquard:2006qi}. Furthermore, we already pointed
out in Ref.~\cite{Marquard:2006qi}, that the result for 
Fig~\ref{fig::selfmaster}$(b)$ (denoted by $I_{16}$ in 
Ref.~\cite{Melnikov:2000zc}) is
wrong. As it turns out, the result given in~\cite{Melnikov:2000zc}  
corresponds to
the integral depicted in Fig.~\ref{fig::selfmaster}$(c)$. Since there is
a one-to-one correspondence between the two integrals with regard to the
IBP relations, it does not
matter which is chosen to be the master integral.

Most of the master integrals of Ref.~\cite{Melnikov:2000zc} were already
calculated in Ref.~\cite{Laporta:1996mq}. However, it turns out that
there are differences in the order $\order{\ep}$ terms of the integrals
$I_2$--$I_7$ of these references. The expressions given in
Ref.~\cite{Melnikov:2000zc} agree with our numerical results.

The results for the renormalization constants can be cast into the
following form
\begin{eqnarray}
  Z_i^{\rm OS} &=& 1 + \frac{\alpha_s(\mu)}{\pi} \left(\frac{e^{\gamma_E}}{4 \pi}
  \right)^{-\epsilon} \delta Z_i^{(1)} +
  \left(\frac{\alpha_s(\mu)}{\pi}\right)^2 \left(\frac{e^{\gamma_E}}{4 \pi}
  \right)^{-2\epsilon} \delta Z_i^{(2)} \nonumber \\
  && + \left(\frac{\alpha_s(\mu)}{\pi}\right)^3
  \left(\frac{e^{\gamma_E}}{4\pi} \right)^{-3\epsilon} \delta Z_i^{(3)}
  + \mathcal{O}\left(\alpha_s^4\right) \,,
  \label{eq::mass::Zm}
\end{eqnarray}
with $i\in\{m,2\}$.
It is convenient to further decompose the three-loop contribution in
terms of the different colour factors
\begin{eqnarray}
  \delta Z_i^{(3)} &=& C_F^3\, Z_i^{FFF} + C_F^2C_A\, Z_i^{FFA} +
  C_FC_A^2\, Z_i^{FAA} \nonumber\\
  && + C_FT_Fn_l \left( C_F\, Z_i^{FFL} + C_A\, Z_i^{FAL} + T_Fn_l\,
  Z_i^{FLL} + T_Fn_h\, Z_i^{FHL} \right) \nonumber\\
  && + C_FT_Fn_h \left( C_F\, Z_i^{FFH} + C_A\, Z_i^{FAH} + T_Fn_h\,
  Z_i^{FHH}
  \right) \,,
  \label{eq::mass::Zm3l}
\end{eqnarray}
where $C_F = (N_c^2 - 1)/(2N_c)$ and $C_A = N_c$ are the eigenvalues of
the quadratic Casimir operators of the fundamental and adjoint
representation of $SU(N_c)$, respectively. In the case of QCD we have
$N_c=3$. $T_F = 1/2$ is the index of the fundamental representation and
$n_f=n_l+n_h$ is the number of quark flavours. $n_l$ and $n_h$ are the
number of light and heavy quark flavours, respectively. The former are
considered to be massless, while the latter have mass $m_q$. Although we
have $n_h=1$ in our applications, we keep a generic label which is useful 
when tracing the origin of the individual contributions. $\alpha_s(\mu)$
is the strong coupling constant defined in the \msbar\ scheme with $n_f$
active flavours.

The one- and two-loop contributions to the mass renormalization constant
are
\begin{eqnarray}
  \delta Z_m^{(1)} &=& - C_F \left[ \frac{3}{4\epsilon} + 1 +
  \frac{3}{4} \Lmu + \left( 2 + \frac{1}{16} \pi^2  + \Lmu + \frac{3}{8}
  \Lmu^2 \right) \epsilon \right. \nonumber \\
  && \left. + \left( 4 + \frac{1}{12} \pi^2 - \frac{1}{4}
  \zeta_3 + \left( 2 + \frac{1}{16}\pi^2 \right) \Lmu + \frac{1}{2}
  \Lmu^2 + \frac{1}{8} \Lmu^3 \right) \epsilon^2 \right]
  \,,
  \label{eq::mass::Zm1l}
\end{eqnarray}
and
\begin{eqnarray}
  \delta Z_m^{(2)} &=&
  \left[ \frac{9}{32 \epsilon^2} + \left( \frac{45}{64} +
  \frac{9}{16} \Lmu \right) \frac{1}{\epsilon} +
  \frac{199}{128} - \frac{17}{64} \pi^2 + \frac{1}{2} \pi^2 \ln 2 -
  \frac{3}{4} \zeta_3 + \frac{45}{32} \Lmu + \frac{9}{16} \Lmu^2
  \right. \nonumber \\
  && + \left( \frac{677}{256} - \frac{205}{128} \pi^2 + 3 \pi^2 \ln 2 -
  \pi^2 \ln^2 2 - \frac{135}{16} \zeta_3 + \frac{7}{40} \pi^4 -
  \frac{1}{2} \ln^4 2 - 12 a_4 \right. \nonumber \\
  && \left.\left. + \left( \frac{199}{64} - \frac{17}{32} \pi^2 +
  \pi^2 \ln 2 - \frac{3}{2} \zeta_3 \right) \Lmu + \frac{45}{32} \Lmu^2
  + \frac{3}{8} \Lmu^3 \right) \epsilon \right] C_F^2 \nonumber \\
  && + \left[ \frac{11}{32 \epsilon^2} - \frac{97}{192 \epsilon} -
  \frac{1111}{384} + \frac{1}{12} \pi^2 - \frac{1}{4} \pi^2 \ln 2 +
  \frac{3}{8} \zeta_3 - \frac{185}{96} \Lmu - \frac{11}{32} \Lmu^2
  \right. \nonumber \\
  && + \left( -\frac{8581}{768} + \frac{271}{1152} \pi^2 - \frac{3}{2}
  \pi^2 \ln 2 + \frac{1}{2} \pi^2 \ln^2 2 + \frac{13}{4} \zeta_3 -
  \frac{7}{80} \pi^4 + \frac{1}{4} \ln^4 2 + 6 a_4 \right. \nonumber \\
  && \left.\left. + \left( -\frac{1463}{192} + \frac{7}{64} \pi^2 -
  \frac{1}{2} \pi^2 \ln 2 + \frac{3}{4} \zeta_3 \right) \Lmu - 
  \frac{229}{96} \Lmu^2 - \frac{11}{32} \Lmu^3\right) \epsilon \right]
  C_AC_F \nonumber \\
  && + \left[ -\frac{1}{8 \epsilon^2} + \frac{5}{48 \epsilon} +
  \frac{71}{96} + \frac{1}{12} \pi^2 + \frac{13}{24} \Lmu + \frac{1}{8}
  \Lmu^2 \right. \nonumber \\
  && \left. + \left( \frac{581}{192} + \frac{97}{288} \pi^2 +
  \zeta_3 + \left( \frac{103}{48} + \frac{3}{16} \pi^2 \right) \Lmu +
  \frac{17}{24} \Lmu^2 + \frac{1}{8} \Lmu^3 \right) \epsilon
  \right] C_FT_Fn_l \nonumber\\
  && + \left[ -\frac{1}{8 \epsilon^2} + \frac{5}{48 \epsilon} +
  \frac{143}{96} - \frac{1}{6} \pi^2 + \frac{13}{24} \Lmu + \frac{1}{8}
  \Lmu^2 + \left( \frac{1133}{192} - \frac{227}{288} \pi^2
  \right. \right. \nonumber \\
  && \left.\left. + \pi^2
  \ln 2 - \frac{7}{2} \zeta_3 + \left( \frac{175}{48} - \frac{5}{16}
  \pi^2 \right) \Lmu + \frac{17}{24} \Lmu^2 + \frac{1}{8} \Lmu^3 \right)
  \epsilon \right] C_FT_Fn_h\,,
  \label{eq::mass::Zm2l}
\end{eqnarray}
where $\Lmu = \ln (\mu^2/M_q^2)$. $\zeta_n$ is Riemann's zeta function
with integer argument $n$ and $a_4={\rm Li}_4(1/2)$. We give the one-
and two-loop results to order $\order{\ep^2}$ and $\order{\ep}$,
respectively. In general these terms are necessary for three-loop calculations.

The individual three-loop terms read
\begin{eqnarray}
  Z_m^{FFF} &=& -\frac{9}{128\ep^3} - \left( \frac{63}{256} +
  \frac{27}{128}\Lmu \right) \frac{1}{\ep^2} + \left( -\frac{457}{512}
  + \frac{111}{512}\pi^2 - \frac{3}{8}\pi^2\ln 2 \right. \nonumber\\
  && \left. + \frac{9}{16}\zeta_3 - \frac{189}{256}\Lmu -
  \frac{81}{256}\Lmu^2 \right) \frac{1}{\ep} - \frac{14225}{3072} -
  \frac{6037}{3072}\pi^2 + 5\pi^2\ln 2 \nonumber\\
  && + \frac{5}{4}\pi^2\ln^2 2 + \frac{153}{128}\zeta_3 -
  \frac{73}{480}\pi^4 - \frac{1}{16}\pi^2\zeta_3 + \frac{5}{8}\zeta_5 -
  \frac{1}{8}\ln^4 2 - 3 a_4 \nonumber\\
  && + \left( -\frac{1371}{512} + \frac{333}{512}\pi^2 -
  \frac{9}{8}\pi^2\ln 2 + \frac{27}{16}\zeta_3 \right) \Lmu -
  \frac{567}{512}\Lmu^2 - \frac{81}{256}\Lmu^3\,,
  \label{eq::mass::zFFF}
\end{eqnarray}
\begin{eqnarray}
  Z_m^{FFA} &=& -\frac{33}{128\ep^3} + \left( \frac{49}{768} -
  \frac{33}{128}\Lmu \right) \frac{1}{\ep^2} + \left( \frac{3311}{1536} -
  \frac{43}{512}\pi^2 + \frac{3}{16}\pi^2\ln 2 \right. \nonumber\\
  && \left. - \frac{9}{32}\zeta_3 + \frac{379}{256}\Lmu +
  \frac{33}{256}\Lmu^2 \right) \frac{1}{\ep} + \frac{100247}{9216} +
  \frac{6545}{9216}\pi^2  + \frac{25}{36}\pi^2\ln 2 \nonumber\\
  && - \frac{89}{72}\pi^2\ln^2 2 - \frac{3995}{384}\zeta_3 +
  \frac{1867}{8640}\pi^4 - \frac{19}{16}\pi^2\zeta_3 +
  \frac{45}{16}\zeta_5 - \frac{35}{144}\ln^4 2 - \frac{35}{6}a_4
  \nonumber\\
  && + \left( \frac{14311}{1536} - \frac{1135}{1536}\pi^2 +
  \frac{71}{48}\pi^2\ln 2 - \frac{71}{32}\zeta_3 \right) \Lmu +
  \frac{1797}{512}\Lmu^2 + \frac{121}{256}\Lmu^3\,,
  \label{eq::mass::zFFA}
\end{eqnarray}
\begin{eqnarray}
  Z_m^{FAA} &=& -\frac{121}{576\ep^3} + \frac{1679}{3456\ep^2} -
  \frac{11413}{20736\ep} - \frac{1322545}{124416} -
  \frac{1955}{3456}\pi^2 - \frac{115}{72}\pi^2\ln 2 \nonumber\\
  && + \frac{11}{36}\pi^2\ln^2 2 + \frac{1343}{288}\zeta_3 -
  \frac{179}{3456}\pi^4 + \frac{51}{64}\pi^2\zeta_3 -
  \frac{65}{32}\zeta_5 + \frac{11}{72}\ln^4 2 + \frac{11}{3} a_4
  \nonumber\\
  && + \left( -\frac{13243}{1728} + \frac{11}{72}\pi^2 -
  \frac{11}{24}\pi^2\ln 2 + \frac{11}{16}\zeta_3 \right) \Lmu -
  \frac{2341}{1152} \Lmu^2 - \frac{121}{576} \Lmu^3\,,
  \label{eq::mass::zFAA}
\end{eqnarray}
\begin{eqnarray}
  Z_m^{FFL} &=& \frac{3}{32 \epsilon^3} + \left( -\frac{5}{192} +
  \frac{3}{32} \Lmu \right) \frac{1}{\epsilon^2} - \left( \frac{65}{384}
  + \frac{7}{128} \pi^2 + \frac{1}{4} \zeta_3 + \frac{23}{64} \Lmu +
  \frac{3}{64} \Lmu^2 \right) \frac{1}{\epsilon} \nonumber \\
  && + \frac{575}{2304} + \frac{1091}{2304} \pi^2 - \frac{11}{9} \pi^2
  \ln 2 + \frac{2}{9} \pi^2 \ln^2 2 + \frac{145}{96} \zeta_3 -
  \frac{119}{2160} \pi^4 + \frac{1}{9} \ln^4 2 \nonumber \\
  && + \frac{8}{3} a_4 - \left( \frac{497}{384} - \frac{5}{384} \pi^2 +
  \frac{1}{3} \pi^2 \ln 2 + \frac{1}{4} \zeta_3 \right) \Lmu -
  \frac{117}{128} \Lmu^2 - \frac{11}{64} \Lmu^3\,,
  \label{eq::mass::zFFL}
\end{eqnarray}
\begin{eqnarray}
  Z_m^{FAL} &=& \frac{11}{72\epsilon^3} - \frac{121}{432 \epsilon^2} +
  \left( \frac{139}{1296} + \frac{1}{4} \zeta_3 \right)
  \frac{1}{\epsilon} + \frac{70763}{15552} + \frac{175}{432} \pi^2 +
  \frac{11}{18} \pi^2 \ln 2 \nonumber \\
  && - \frac{1}{9} \pi^2 \ln^2 2 + \frac{89}{144} \zeta_3 +
  \frac{19}{2160} \pi^4  - \frac{1}{18} \ln^4 2 - \frac{4}{3} a_4
  \nonumber \\
  && + \left( \frac{869}{216} + \frac{7}{72} \pi^2 + \frac{1}{6} \pi^2
  \ln 2 + \frac{1}{2} \zeta_3 \right) \Lmu + \frac{373}{288} \Lmu^2 +
  \frac{11}{72} \Lmu^3\,,
  \label{eq::mass::zFAL}
\end{eqnarray}
\begin{eqnarray}
  Z_m^{FLL} &=& -\frac{1}{36 \epsilon^3} + \frac{5}{216 \epsilon^2} +
  \frac{35}{1296 \epsilon} - \frac{2353}{7776} - \frac{13}{108} \pi^2 -
  \frac{7}{18} \zeta_3 \nonumber \\
  && - \left( \frac{89}{216} + \frac{1}{18} \pi^2 \right)
  \Lmu - \frac{13}{72} \Lmu^2 - \frac{1}{36} \Lmu^3\,,
  \label{eq::mass::zFLL}
\end{eqnarray}
\begin{eqnarray}
  Z_m^{FHL} &=& -\frac{1}{18 \epsilon^3} + \frac{5}{108 \epsilon^2} +
  \frac{35}{648 \epsilon} - \frac{5917}{3888} + \frac{13}{108} \pi^2 +
  \frac{2}{9} \zeta_3 \nonumber \\
  && - \left( \frac{143}{108} - \frac{1}{18} \pi^2 \right)
  \Lmu - \frac{13}{36} \Lmu^2 - \frac{1}{18} \Lmu^3\,,
  \label{eq::mass::zFHL}
\end{eqnarray}
\begin{eqnarray}
  Z_m^{FFH} &=& \frac{3}{32 \epsilon^3} + \left( -\frac{5}{192} +
  \frac{3}{32} \Lmu \right) \frac{1}{\epsilon^2} - \left( \frac{281}{384}
  - \frac{17}{128} \pi^2 + \frac{1}{4} \zeta_3 + \frac{23}{64} \Lmu +
  \frac{3}{64} \Lmu^2 \right) \frac{1}{\epsilon} \nonumber \\
  && - \frac{5257}{2304} - \frac{1327}{6912} \pi^2 + \frac{5}{36} \pi^2
  \ln 2 - \frac{1}{9} \pi^2 \ln^2 2 + \frac{37}{96} \zeta_3 +
  \frac{91}{2160} \pi^4 + \frac{1}{9} \ln^4 2 \nonumber \\
  && + \frac{8}{3} a_4 - \left( \frac{1145}{384} - \frac{221}{384} \pi^2 +
  \frac{1}{3} \pi^2 \ln 2 + \frac{1}{4} \zeta_3 \right) \Lmu -
  \frac{117}{128} \Lmu^2 - \frac{11}{64} \Lmu^3\,,
  \label{eq::mass::zFFH}
\end{eqnarray}
\begin{eqnarray}
  Z_m^{FAH} &=& \frac{11}{72\epsilon^3} - \frac{121}{432 \epsilon^2} +
  \left( \frac{139}{1296} + \frac{1}{4} \zeta_3 \right)
  \frac{1}{\epsilon} + \frac{144959}{15552} - \frac{449}{144} \pi^2 +
  \frac{32}{9} \pi^2 \ln 2 \nonumber \\
  && + \frac{1}{18} \pi^2 \ln^2 2 - \frac{109}{144} \zeta_3 -
  \frac{43}{1080} \pi^4 + \frac{1}{8}\pi^2\zeta_3 - \frac{5}{8} \zeta_5
  - \frac{1}{18} \ln^4 2 - \frac{4}{3} a_4 \nonumber \\
  && + \left( \frac{583}{108} - \frac{13}{36} \pi^2 + \frac{1}{6} \pi^2
  \ln 2 + \frac{1}{2} \zeta_3 \right) \Lmu + \frac{373}{288} \Lmu^2 +
  \frac{11}{72} \Lmu^3\,,
  \label{eq::mass::zFAH}
\end{eqnarray}
\begin{eqnarray}
  Z_m^{FHH} &=& -\frac{1}{36 \epsilon^3} + \frac{5}{216 \epsilon^2} +
  \frac{35}{1296 \epsilon} - \frac{9481}{7776} + \frac{4}{135} \pi^2 +
  \frac{11}{18} \zeta_3 \nonumber \\
  && - \left( \frac{197}{216} - \frac{1}{9} \pi^2 \right)
  \Lmu - \frac{13}{72} \Lmu^2 - \frac{1}{36} \Lmu^3\,.
  \label{eq::mass::zFHH}
\end{eqnarray}

For the wave function renormalization constant, we have at the one-loop
level $\delta Z_2^{(1)}=\delta Z_m^{(1)}$. The two-loop contribution to
the wave function renormalization constant reads
\begin{eqnarray}
  \delta Z_2^{(2)} &=&
  \left[ \frac{9}{32 \epsilon^2} + \left( \frac{51}{64} +
  \frac{9}{16} \Lmu \right) \frac{1}{\epsilon} +
  \frac{433}{128} - \frac{49}{64} \pi^2 + \pi^2 \ln 2 -
  \frac{3}{2} \zeta_3 + \frac{51}{32} \Lmu + \frac{9}{16} \Lmu^2
  \right. \nonumber \\
  && + \left( \frac{211}{256} - \frac{339}{128} \pi^2 +
  \frac{23}{4} \pi^2 \ln 2 - 2 \pi^2 \ln^2 2 - \frac{297}{16} \zeta_3 +
  \frac{7}{20} \pi^4 - \ln^4 2 - 24 a_4 \right. \nonumber \\
  && \left.\left. + \left( \frac{433}{64} - \frac{49}{32} \pi^2 + 2
  \pi^2 \ln 2 - 3 \zeta_3 \right) \Lmu + \frac{51}{32} \Lmu^2 +
  \frac{3}{8} \Lmu^3 \right) \epsilon \right] C_F^2 \nonumber \\
  && + \left[ \frac{11}{32
  \epsilon^2} - \frac{127}{192 \epsilon} - \frac{1705}{384} +
  \frac{5}{16} \pi^2 - \frac{1}{2} \pi^2 \ln 2 + \frac{3}{4} \zeta_3
  - \frac{215}{96} \Lmu - \frac{11}{32} \Lmu^2 \right. \nonumber \\
  && + \left( -\frac{9907}{768} + \frac{769}{1152} \pi^2 - \frac{23}{8}
  \pi^2 \ln 2 + \pi^2 \ln^2 2 + \frac{129}{16} \zeta_3 - \frac{7}{40}
  \pi^4 + \frac{1}{2} \ln^4 2 + 12 a_4 \right. \nonumber \\
  && \left.\left. + \left( -\frac{2057}{192} +
  \frac{109}{192} \pi^2 - \pi^2 \ln 2 + \frac{3}{2} \zeta_3 \right) \Lmu -
  \frac{259}{96} \Lmu^2 - \frac{11}{32} \Lmu^3\right) \epsilon \right]
  C_AC_F \nonumber \\
  && + \left[ -\frac{1}{8 \epsilon^2} + \frac{11}{48 \epsilon} +
  \frac{113}{96} + \frac{1}{12} \pi^2 + \frac{19}{24} \Lmu + \frac{1}{8}
  \Lmu^2 \right. \nonumber \\
  && \left. + \left( \frac{851}{192} + \frac{127}{288} \pi^2 +
  \zeta_3 + \left( \frac{145}{48} + \frac{3}{16} \pi^2 \right) \Lmu +
  \frac{23}{24} \Lmu^2 + \frac{1}{8} \Lmu^3 \right) \epsilon
  \right] C_FT_Fn_l \nonumber\\
  && + \left[ \left( \frac{1}{16} + \frac{1}{4} \Lmu \right)
  \frac{1}{\epsilon} + \frac{947}{288} - \frac{5}{16} \pi^2
  +\frac{11}{24} \Lmu + \frac{3}{8} \Lmu^2 + \left( \frac{17971}{1728} -
  \frac{445}{288} \pi^2 \right.\right. \nonumber \\ 
  && \left.\left. + 2
  \pi^2 \ln 2 - \frac{85}{12} \zeta_3 + \left( \frac{1043}{144} -
  \frac{29}{48} \pi^2 \right) \Lmu + \frac{5}{8} \Lmu^2 + \frac{7}{24}
  \Lmu^3 \right) \epsilon \right] C_FT_Fn_h\,.
  \label{eq::wave::Z22l}
\end{eqnarray}
Again, we give the result to order $\order{\ep}$, which is necessary for
three-loop calculations.

The individual three-loop terms are given by
\begin{eqnarray}
  Z_2^{FFF} &=& -\frac{9}{128\ep^3} - \left( \frac{81}{256} +
  \frac{27}{128}\Lmu \right) \frac{1}{\ep^2} + \left( -\frac{1039}{512}
  + \frac{303}{512}\pi^2 - \frac{3}{4}\pi^2\ln 2 \right. \nonumber\\
  && \left. + \frac{9}{8}\zeta_3 - \frac{243}{256}\Lmu -
  \frac{81}{256}\Lmu^2 \right) \frac{1}{\ep} - \frac{10823}{3072} -
  \frac{58321}{9216}\pi^2 + \frac{685}{48}\pi^2\ln 2 \nonumber\\
  && + 3\pi^2\ln^2 2 - \frac{739}{128}\zeta_3 - \frac{41}{120}\pi^4 +
  \frac{1}{8}\pi^2\zeta_3 - \frac{5}{16}\zeta_5 - \frac{5}{12}\ln^4 2 -
  10a_4 \nonumber\\
  && + \left( -\frac{3117}{512} + \frac{909}{512}\pi^2 -
  \frac{9}{4}\pi^2\ln 2 + \frac{27}{8}\zeta_3 \right) \Lmu -
  \frac{729}{512}\Lmu^2 - \frac{81}{256}\Lmu^3\,,
  \label{eq::wave::zFFF}
\end{eqnarray}
\begin{eqnarray}
  Z_2^{FFA} &=& -\frac{33}{128\ep^3} + \left( \frac{95}{768} -
  \frac{33}{128}\Lmu \right) \frac{1}{\ep^2} + \left( \frac{1787}{512} -
  \frac{131}{512}\pi^2 + \frac{3}{8}\pi^2\ln 2 \right. \nonumber\\
  && \left. - \frac{5}{8}\zeta_3 + \frac{469}{256}\Lmu +
  \frac{33}{256}\Lmu^2 \right) \frac{1}{\ep} + \frac{136945}{9216} +
  \frac{29695}{9216}\pi^2  - \frac{755}{288}\pi^2\ln 2 \nonumber\\
  && - \frac{235}{72}\pi^2\ln^2 2 - \frac{6913}{384}\zeta_3 +
  \frac{1793}{3456}\pi^4 - \frac{45}{16}\pi^2\zeta_3 +
  \frac{145}{16}\zeta_5 - \frac{55}{144}\ln^4 2 - \frac{55}{6}a_4
  \nonumber\\
  && + \left( \frac{25609}{1536} - \frac{3335}{1536}\pi^2 +
  \frac{71}{24}\pi^2\ln 2 - \frac{37}{8}\zeta_3 \right) \Lmu +
  \frac{2155}{512}\Lmu^2 + \frac{121}{256}\Lmu^3\,,
  \label{eq::wave::zFFA}
\end{eqnarray}
\begin{eqnarray}
  Z_2^{FAA} &=& -\frac{121}{576\ep^3} + \frac{2009}{3456\ep^2} - \left[
  \frac{12793}{20736} + \frac{3}{128}\zeta_3 + \frac{1}{1080}\pi^4 +
  \left( \frac{1}{768} + \frac{3}{256}\zeta_3 \right.\right. \nonumber\\
  && \left.\left. -\frac{1}{4320}\pi^4 \right)\xi \right] \frac{1}{\ep}
  - \frac{1654711}{124416} - \frac{4339}{3456}\pi^2 -
  \frac{325}{144}\pi^2\ln 2 + \frac{127}{144}\pi^2\ln^2 2 \nonumber\\
  && + \frac{5857}{576}\zeta_3 - \frac{3419}{23040}\pi^4 +
  \frac{127}{72}\pi^2\zeta_3 - \frac{37}{6}\zeta_5 + \frac{85}{288}\ln^4
  2 + \frac{85}{12}a_4 \nonumber\\
  && + \left( - \frac{13}{768} - \frac{1}{256}\pi^2 -
  \frac{13}{256}\zeta_3 + \frac{17}{27648}\pi^4 +
  \frac{1}{144}\pi^2\zeta_3 + \frac{7}{384}\zeta_5 \right)\xi
  \nonumber\\
  && + \left[ - \frac{36977}{3456} + \frac{55}{96}\pi^2 -
  \frac{11}{12}\pi^2\ln 2 + \frac{167}{128}\zeta_3 - \frac{1}{360}\pi^4
  \right. \nonumber\\
  && \left. + \left( - \frac{1}{256} - \frac{9}{256}\zeta_3 +
  \frac{1}{1440}\pi^4 \right)\xi \right] \Lmu - \frac{2671}{1152} \Lmu^2
  - \frac{121}{576} \Lmu^3\,,
  \label{eq::wave::zFAA}
\end{eqnarray}
\begin{eqnarray}
  Z_2^{FFL} &=& \frac{3}{32 \epsilon^3} + \left( -\frac{19}{192} +
  \frac{3}{32} \Lmu \right) \frac{1}{\epsilon^2} - \left(
  \frac{235}{384} + \frac{7}{128} \pi^2 + \frac{1}{4} \zeta_3 +
  \frac{41}{64} \Lmu + \frac{3}{64} \Lmu^2 \right) \frac{1}{\epsilon}
  \nonumber \\
  && -\frac{3083}{2304} + \frac{2845}{2304} \pi^2 -
  \frac{47}{18} \pi^2 \ln 2 + \frac{4}{9} \pi^2 \ln^2 2 +
  \frac{473}{96} \zeta_3 - \frac{229}{2160} \pi^4 + \frac{2}{9}
  \ln^4 2 \nonumber \\
  && + \frac{16}{3} a_4 + \left( -\frac{1475}{384} + \frac{133}{384}
  \pi^2 - \frac{2}{3} \pi^2 \ln 2 + \frac{1}{4} \zeta_3 \right) \Lmu -
  \frac{179}{128} \Lmu^2 - \frac{11}{64} \Lmu^3\,,
  \label{eq::wave::zFFL}
\end{eqnarray}
\begin{eqnarray}
  Z_2^{FAL} &=&  \frac{11}{72\epsilon^3} - \frac{169}{432 \epsilon^2} +
  \left( \frac{313}{1296} + \frac{1}{4} \zeta_3 \right)
  \frac{1}{\epsilon} + \frac{111791}{15552} + \frac{13}{48} \pi^2 +
  \frac{47}{36} \pi^2 \ln 2 \nonumber \\
  &&- \frac{2}{9} \pi^2 \ln^2 2 - \frac{35}{72} \zeta_3 +
  \frac{19}{1080} \pi^4  - \frac{1}{9} \ln^4 2 - \frac{8}{3} a_4
  \nonumber \\
  && + \left( \frac{169}{27} - \frac{1}{18} \pi^2 + \frac{1}{3} \pi^2
  \ln 2 + \frac{1}{4} \zeta_3 \right) \Lmu + \frac{469}{288} \Lmu^2 +
  \frac{11}{72} \Lmu^3\,,
  \label{eq::wave::zFAL}
\end{eqnarray}
\begin{eqnarray}
  Z_2^{FLL} &=& -\frac{1}{36 \epsilon^3} + \frac{11}{216 \epsilon^2} +
  \frac{5}{1296 \epsilon} - \frac{5767}{7776} - \frac{19}{108} \pi^2 -
  \frac{7}{18} \zeta_3 \nonumber \\
  && - \left( \frac{167}{216} + \frac{1}{18} \pi^2 \right)
  \Lmu - \frac{19}{72} \Lmu^2 - \frac{1}{36} \Lmu^3\,,
  \label{eq::wave::zFLL}
\end{eqnarray}
\begin{eqnarray}
  Z_2^{FHL} &=& \left( \frac{1}{36} + \frac{1}{12} \Lmu \right)
  \frac{1}{\epsilon^2} + \left( -\frac{5}{216} + \frac{1}{144} \pi^2 -
  \frac{1}{9} \Lmu + \frac{1}{24} \Lmu^2 \right) \frac{1}{\epsilon}
  \nonumber \\
  && - \frac{4721}{1296} + \frac{19}{54} \pi^2 - \frac{1}{36}
  \zeta_3 + \left( -\frac{329}{108} + \frac{25}{144} \pi^2 \right)
  \Lmu - \frac{7}{12} \Lmu^2 - \frac{5}{72} \Lmu^3\,,
  \label{eq::wave::zFHL}
\end{eqnarray}
\begin{eqnarray}
  Z_2^{FFH} &=& -\left( \frac{7}{192} + \frac{3}{16}\Lmu \right)
  \frac{1}{\ep^2} - \left(\frac{707}{384} - \frac{15}{64}\pi^2 +
  \frac{29}{64}\Lmu + \frac{15}{32}\Lmu^2 \right) \frac{1}{\ep}
  \nonumber\\
  && - \frac{76897}{6912} - \frac{11551}{20736}\pi^2 +
  \frac{7}{18}\pi^2\ln 2 - \frac{1}{2}\pi^2\ln^2 2 +
  \frac{1763}{288}\zeta_3 + \frac{31}{720}\pi^4 + \frac{1}{2}\ln^4 2
  \nonumber\\
  && + 12a_4 + \left( -\frac{2891}{384} + \frac{233}{192}\pi^2 -
  \frac{2}{3}\pi^2\ln 2 + \zeta_3 \right) \Lmu - \frac{143}{128} \Lmu^2
  - \frac{19}{32} \Lmu^3\,,
  \label{eq::wave::zFFH}
\end{eqnarray}
\begin{eqnarray}
  Z_2^{FAH} &=& \frac{1 - \xi}{192\ep^3} - \left[ \frac{7}{72} -
  \frac{1}{64}\xi + \left( \frac{41}{192} + \frac{1}{64}\xi \right) \Lmu
  \right] \frac{1}{\ep^2} + \left[ \frac{13}{216} - \frac{41}{2304}\pi^2
  \right. \nonumber\\
  && \left. - \left( \frac{35}{576} + \frac{1}{768}\pi^2 \right)\xi +
  \left( \frac{83}{144} + \frac{3}{64}\xi \right) \Lmu - \left(
  \frac{35}{384} + \frac{3}{128}\xi \right) \Lmu^2 \right] \frac{1}{\ep}
  \nonumber\\
  && + \frac{49901}{2592} - \frac{36019}{5184}\pi^2 +
  \frac{80}{9}\pi^2\ln 2 + \frac{1}{3}\pi^2\ln^2 2 -
  \frac{77}{16}\zeta_3 - \frac{17}{360}\pi^4 + \frac{11}{48}\pi^2\zeta_3
  \nonumber\\
  && - \frac{15}{16}\zeta_5 - \frac{1}{3}\ln^4 2 - 8a_4 + \left(
  \frac{407}{1728} + \frac{1}{256}\pi^2 - \frac{7}{192}\zeta_3
  \right)\xi \nonumber\\
  && + \left[ \frac{4141}{432} - \frac{641}{768}\pi^2 +
  \frac{1}{3}\pi^2\ln 2 - \frac{1}{2}\zeta_3 - \left( \frac{35}{192} +
  \frac{1}{256}\pi^2 \right)\xi \right] \Lmu \nonumber\\
  && + \left( \frac{35}{16} + \frac{9}{128}\xi \right) \Lmu^2 + \left(
  \frac{247}{1152} - \frac{3}{128}\xi \right) \Lmu^3\,,
  \label{eq::wave::zFAH}
\end{eqnarray}
\begin{eqnarray}
  Z_2^{FHH} &=& \frac{1}{72\ep^2} - \left( \frac{5}{432} + \frac{1}{12}
  \Lmu^2 \right) \frac{1}{\ep} - \frac{8425}{2592} + \frac{2}{45}\pi^2 +
  \frac{7}{3}\zeta_3 \nonumber\\
  && - \left( \frac{481}{216} - \frac{5}{24}\pi^2 \right) \Lmu -
  \frac{11}{72} \Lmu^2 - \frac{1}{6} \Lmu^3\,.
  \label{eq::wave::zFHH}
\end{eqnarray}
Starting from the three-loop level, the wave function renormalization
constant depends on the gauge parameter, $\xi$. The parameter in the
above equations is defined through the gluon propagator as
\begin{equation}
  D_{\mu\nu}^{ab}(k) = -\frac{i}{k^2}\, \left( g_{\mu\nu} - \xi\,
  \frac{k_\mu k_\nu}{k^2} \right)\, \delta^{ab}\,,
  \label{eq::wave::gluon}
\end{equation}
where $a$ and $b$ are colour indices.

We want to mention that our results for $Z_m^{\rm OS}$ and
$Z_2^{\rm OS}$ agree with the
literature~\cite{Melnikov:2000qh,Melnikov:2000zc}. Whereas
in~\cite{Melnikov:2000qh,Melnikov:2000zc} they
are expressed in terms of the bare coupling we decided
to use the renormalized $\alpha_s$ as an expansion parameter which to
our opinion is more convenient in practical applications.

The genuine three-loop integrals which appear in the \dreg{}
calculation are the same for \dred{}. The main complication is the
more involved renormalization which is discussed in more detail in the
next Section.


\section{\label{sec::ren_dred}Renormalization in \dred{}}

Let us in this Section collect the \dred{} counterterms 
needed for our calculation.
In addition to the strong coupling $\alpha_s$ also the evanescent 
coupling\footnote{We refer to~\cite{Jack:1993ws,Harlander:2006xq} for a precise
  definition of the evanescent couplings.}
$\alpha_e$ has to be renormalized to two-loop order.
The evanescent 
couplings\footnote{ For
  the $SU(3)$ gauge group, which we exclusively consider in this
  paper,
  there are three independent such couplings~\cite{Jack:1993ws}.} 
$\eta_1$, $\eta_2$ and $\eta_3$ appear for the 
first time at three-loop order and thus no renormalization is
necessary. 
Both for  the heavy quark mass, $m_q$, and the $\varepsilon$-scalar mass,
$m_\varepsilon$, two-loop counterterms are necessary.
Whereas the couplings are renormalized using minimal subtraction the
masses are renormalized on-shell. The corresponding counterterms are
defined through
\begin{align}
  \alpha_s^{0,\overline{\rm DR}} &= 
  \mu^{2\epsilon} \left(Z_s^{\overline{\rm DR}}\right)^2 
  \alpha_s^{\overline{\rm DR}}\,,\qquad &
  \alpha_e^0   &= 
  \mu^{2\epsilon} \left(Z_e\right)^2 \alpha_e\,,\qquad &
  \nonumber\\
  m_q^{0,\overline{\rm DR}} &= 
  M_q Z_m^{\rm OS,\overline{\rm DR}}\,,\qquad &
  \left(m_\varepsilon^0\right)^2 &= 
  m_\varepsilon^2 Z_{m_\varepsilon}^{\rm OS}\,.\qquad &
  \label{eq::renconst}  
\end{align}
We attach an additional index ``$\overline{\rm DR}$'' to the quark mass
renormalization constant in order to remind that it relates the pole
mass to the bare mass in the \dred{} 
scheme\footnote{In principle such an index would also be necessary in
  Section~\ref{sec::dimreg}. However, since the $\overline{\rm MS}$
  scheme in connection with $\dreg{}$ constitutes the standard
  framework we refrain from introducing an additional index there.} 
(in this context see also Ref.~\cite{Harlander:2006rj}).

Recently the quantities $Z_e$ and $Z_s^{\overline{\rm DR}}$ have been computed to
three- and four-loop order~\cite{Harlander:2006rj,Harlander:2006xq},
respectively. The results 
have been presented in terms of the corresponding $\beta$ functions.
For completeness we present in the following the two-loop results for the
renormalization constants
\begin{eqnarray}
  Z_s^{\overline{\rm DR}} &=& 1 + 
  \apiDR\frac{1}{\ep}\left(-\frac{11}{24} C_A + \frac{1}{6}
  T_F n_f\right)
  +\left(\apiDR \right)^2\bigg[
    \frac{1}{\ep^2}\left( \frac{121}{384} C_A^2 - \frac{11}{48} C_A
    T_F  n_f
    \right.\nonumber\\&&\left.\mbox{}
    + \frac{1}{24} n_f^2 T_F^2\right) 
    + \frac{1}{\ep}\left(-\frac{17}{96} C_A^2 +
    \frac{5}{48} C_A T_F n_f + \frac{1}{16} C_F T_F n_f\right)  
    \bigg]
  \,,
  \\
  Z_e &=& 1 + \apiDR \frac{1}{\ep} \left(-\frac{3}{4} C_F\right) 
  + \aepi
  \frac{1}{\ep} \left(-\frac{1}{4} C_A + \frac{1}{2} C_F  +
  \frac{1}{4} T_F n_f \right)
  \nonumber\\&&\mbox{} 
  + \left(\apiDR \right)^2\bigg[
    \frac{1}{\ep^2}\left(\frac{11}{32} C_A C_F 
    + \frac{9}{32} C_F^2  -
    \frac{1}{8}  C_F T_F n_f  \right) 
    +\frac{1}{\ep}\left( \frac{7}{256} C_A^2 - \frac{55}{192} C_A C_F
    \right.\nonumber\\&&\left.\mbox{} 
    - \frac{3}{64} C_F^2-\frac{1}{32} C_A T_F n_f  +
    \frac{5}{48} C_F T_F n_f  \right)
    \bigg] 
  + \apiDR \aepi \bigg[
    \frac{1}{\ep^2} \left(\frac{3}{8} C_A C_F  - \frac{3}{4} C_F^2
    \right.\nonumber\\&&\left.\mbox{} 
    - \frac{3}{8}  C_F T_F n_f  \right) 
    +\frac{1}{\ep} \left( \frac{3}{32} C_A^2 - \frac{5}{8} C_A C_F 
    + \frac{11}{16} C_F^2  + \frac{5}{32} C_F T_F n_f  \right)
    \bigg]
  \nonumber\\&&\mbox{}
  + \left(\aepi\right)^2\bigg[
    \frac{1}{\ep^2} \left(\frac{3}{32} C_A^2  - \frac{3}{8} C_A C_F 
    + \frac{3}{8} C_F^2  - \frac{3}{16} C_A T_F n_f  + 
    \frac{3}{8} C_F T_F n_f +\frac{3}{32} T_F^2 n_f^2
    \right)
    \nonumber\\&&\mbox{} 
    +\frac{1}{\ep} \left( -\frac{3}{32} C_A^2 + \frac{5}{16} C_A C_F 
    -\frac{1}{4} C_F^2  + \frac{3}{32} C_A T_F n_f -\frac{3}{16} C_F
    T_F n_f \right)
    \bigg] 
  \nonumber\\&&\mbox{} 
  +  \aepi \frac{1}{\ep} \left(\frac{\eta_1}{\pi} \frac{9}{32}
    -\frac{\eta_2}{\pi} 
  \frac{5}{16} -\frac{\eta_3}{\pi} \frac{3}{16}\right)
  -\left(\frac{\eta_1}{\pi} \right)^2 \frac{1}{\ep} \frac{27}{256} 
  + \left(\frac{\eta_2}{\pi} \right)^2\frac{1}{\ep} \frac{15}{16} 
  +\frac{\eta_1}{\pi} \frac{\eta_3}{\pi}
  \frac{1}{\ep} 
  \frac{9}{64} 
  \nonumber\\&&\mbox{} 
  - \left(\frac{\eta_3}{\pi} \right)^2 \frac{1}{\ep}
  \frac{21}{128} 
  \,.
\end{eqnarray}
Note that we set $N_c=3$ in all terms containing the couplings
$\eta_i$ since our implementation of these couplings is only valid for
$SU(3)$. We would also like to point out that the analytical form of
$Z_s^{\overline{\rm DR}}$ is identical to the corresponding result in
the $\overline{\rm MS}$ scheme. This has been shown by an explicit
calculation in Ref.~\cite{Capper:1979ns}. The one-loop result for $Z_e$
can be found in Ref.~\cite{Jack:1993ws}.

The one-loop corrections to $Z_m^{\rm OS,\overline{\rm DR}}$ can be found in
Ref.~\cite{Martin:1993yx}
and the two-loop terms have been computed in 
Ref.~\cite{Avdeev:1997sz}. For our calculation we also need the ${\cal
  O}(\ep^2)$ 
and ${\cal O}(\ep)$ parts of the one- and two-loop terms,
respectively.

In Section~\ref{sec::dimred} we want to 
present the finite result obtained by considering the ratio
$Z_m^{\rm OS, \overline{\rm DR}}/Z_m^{\overline{\rm DR}}$.
The quantity $Z_m^{\overline{\rm DR}}$ has been computed in
Ref.~\cite{Harlander:2006rj} to three and in
Ref.~\cite{Harlander:2006xq} even to four-loop 
order. Whereas in~\cite{Harlander:2006rj,Harlander:2006xq} only the
anomalous dimensions are 
given we want to present the explicit result for the renormalization
constant 
\begin{eqnarray}
  Z_m^{\overline{\rm DR}} &=& 1 + \apiDR  \frac{1}{\ep} \left(-\frac{3}{4}
  C_F \right) 
  + \left(\apiDR\right)^2\bigg[
    \frac{1}{\ep^2} \left(\frac{11}{32} C_A C_F + \frac{9}{32} C_F^2 
    -\frac{1}{8} C_F T_F n_f  \right)
    \nonumber\\&&\mbox{}
    +\frac{1}{\ep}\left(-\frac{91}{192} C_A C_F - \frac{3}{64} C_F^2 
    + \frac{5}{48} C_F T_F n_f \right)
    \bigg]
  + \apiDR \aepi
  \left(\frac{3}{16}\frac{1}{\ep} C_F^2\right)
  \nonumber\\&&\mbox{}
  + \left(\aepi\right)^2 \frac{1}{\ep}\left(
  \frac{1}{16} C_A C_F - \frac{1}{8} C_F^2  - \frac{1}{16} C_F T_F n_f
  \right)
  +\left(\apiDR\right)^3\bigg[
    \frac{1}{\ep^3} \bigg(
    -\frac{121}{576} C_A^2 C_F  
    \nonumber\\&&\mbox{}
    - \frac{33}{128} C_A C_F^2 
    - \frac{9}{128} C_F^3  
    + \frac{11}{72} C_A C_F T_F n_f
    + \frac{3}{32} C_F^2 T_F n_f - \frac{1}{36} C_F T_F^2 n_f^2
    \bigg)
    \nonumber\\&&\mbox{}
    +\frac{1}{\ep^2} \bigg(
    \frac{1613}{3456}C_A^2 C_F + \frac{295}{768} C_A C_F^2 
    + \frac{9}{256} C_F^3 
    -\frac{59}{216} C_A C_F T_F n_f - \frac{29}{192} C_F^2 T_F n_f
    \nonumber\\&&\mbox{}
    + \frac{5}{216} C_F T_F^2 n_f^2 
    \bigg)
    +\frac{1}{\ep} \bigg(
    -\frac{10255}{20736} C_A^2 C_F + \frac{133}{768} C_A C_F^2 
    - \frac{43}{128} C_F^3 
    +\left(\frac{281}{2592}
    \right.\nonumber\\&&\left.\mbox{}
    + \frac{1}{4} \zeta_3\right) C_A C_F T_F n_f 
    + \left(\frac{23}{96}- \frac{1}{4} \zeta_3 \right)C_F^2 T_F n_f 
    + \frac{35}{1296} C_F  T_F^2 n_f^2
    \bigg)
    \bigg]
  \nonumber\\&&\mbox{}
  +\left(\apiDR\right)^2\aepi\bigg[
    \frac{1}{\ep^2} \left(
    -\frac{11}{192} C_A C_F^2 - \frac{15}{64} C_F^3 + \frac{1}{48} C_F^2
    T_F n_f \right)
    +\frac{1}{\ep}\left(
    \frac{5}{256} C_A^2 C_F 
    \right.\nonumber\\&&\left.\mbox{}
    + \frac{7}{32} C_A C_F^2 + \frac{9}{64} C_F^3
    - \frac{3}{32} C_F^2 T_F n_f \right)
    \bigg]
  +\apiDR \left(\aepi\right)^2 \bigg[
    \frac{1}{\ep^2} \left(
    -\frac{9}{64} C_A C_F^2 
    + \frac{9}{32} C_F^3  
    \right.\nonumber\\&&\left.\mbox{}
    + \frac{9}{64} C_F^2
    T_F n_f \right)
    +\frac{1}{\ep} \left(
    -\frac{1}{64} C_A^2 C_F  + \frac{7}{32} C_A C_F^2  - \frac{3}{8} C_F^3
    - \frac{1}{64} C_A C_F T_F n_f
    \right.\nonumber\\&&\left.\mbox{}
    - \frac{1}{8} C_F^2 T_F n_f 
    \right)
    \bigg]
  +\left(\aepi\right)^3 \bigg[
    \frac{1}{\ep^2} \bigg(
    -\frac{1}{48} C_A^2 C_F 
    + \frac{1}{12} C_A C_F^2 
    -\frac{1}{12} C_F^3  
    \nonumber\\&&\mbox{}
    + \frac{1}{24} C_A C_F T_F n_f 
    -\frac{1}{12} C_F^2 T_F n_f  - \frac{1}{48} C_F T_F^2 n_f^2 
    \bigg)
    +\frac{1}{\ep} \bigg(
    \frac{1}{32} C_A^2 C_F - \frac{1}{8} C_A C_F^2 
    + \frac{1}{8} C_F^3  
    \nonumber\\&&\mbox{}
    - \frac{1}{24} C_A C_F T_F n_f 
    +\frac{5}{48} C_F^2 T_F n_f + \frac{1}{96} C_F T_F^2 n_f^2 
    \bigg)
    \bigg]
  - \frac{1}{8} \left(\aepi\right)^2\frac{\eta_1}{\pi}\frac{1}{\ep}
  \nonumber\\&&
  +\frac{5}{36} \left(\aepi\right)^2\frac{\eta_2}{\pi}\frac{1}{\ep}
  +\frac{1}{12} \left(\aepi\right)^2\frac{\eta_3}{\pi}\frac{1}{\ep}
  +\frac{3}{64} \aepi \left(\frac{\eta_1}{\pi}\right)^2\frac{1}{\ep}
  -\frac{5}{12} \aepi \left(\frac{\eta_2}{\pi}\right)^2\frac{1}{\ep}
  \nonumber\\&&
  +\frac{7}{96} \aepi \left(\frac{\eta_3}{\pi}\right)^2\frac{1}{\ep}
  -\frac{1}{16} \aepi\frac{\eta_1}{\pi}\frac{\eta_3}{\pi}\frac{1}{\ep}
  \,.
  \label{eq::ZmDR}
\end{eqnarray}

\begin{figure}[t]
  \begin{center}
    \begin{tabular}{c}
      \leavevmode
      \epsfxsize=.95\textwidth
      \epsffile{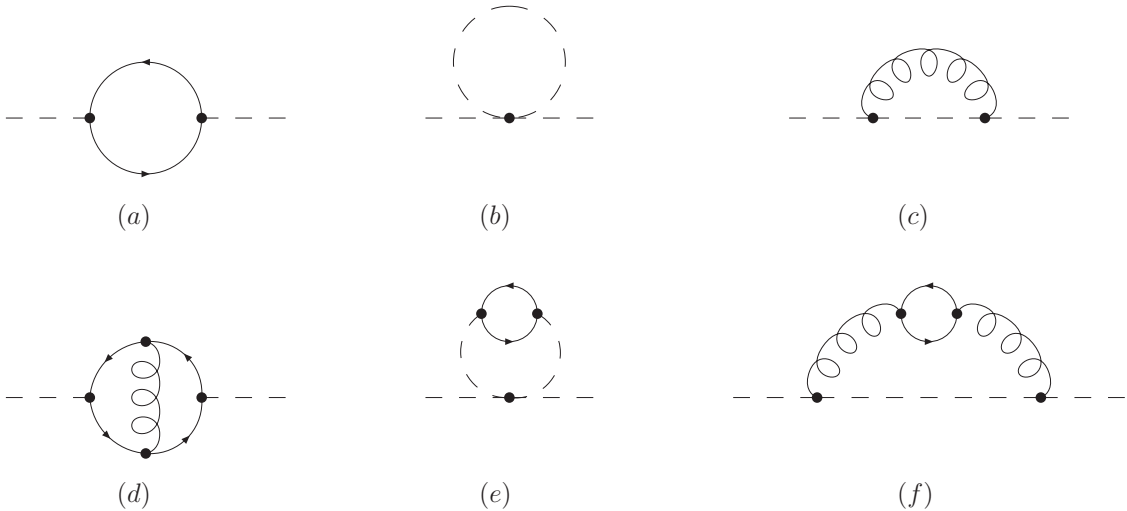}
    \end{tabular}
  \end{center}
  \caption{\label{fig::eps_prop}
    One- and two-loop Feynman diagrams contributing to the 
    $\varepsilon$-scalar propagator. Dashed lines denote
    $\varepsilon$ scalars, curly lines denote gluons and solid lines
    denote massive quarks with mass $m_q$.
          }
\end{figure}

In order to achieve the
 finite result for the relation between the
pole and the \drbar{} quark mass it is necessary to fix a
renormalization scheme also for the mass of the $\varepsilon$ scalar,
$m_\varepsilon$. Although,
there is in general no tree-level term in the Lagrangian, 
 there are loop induced  contributions to
$m_\varepsilon$  which require the introduction of corresponding
counterterms. The relevant Feynman diagrams contributing to the 
$\varepsilon$-scalar propagator show  quadratic divergences and
 therefore, one  needs to consider 
only contributions from massive particles. Thus, in our case, only
diagrams involving a massive quark have to be taken into account. Some
 sample diagrams are 
shown in Fig.~\ref{fig::eps_prop}.

It is common practice to renormalize $m_\varepsilon$ on-shell and
require that the renormalized mass is zero to each order in
perturbation theory~\cite{Jack:1994rk}. This scheme is known as the 
\drbar{}$^\prime$ scheme~\cite{Jack:1994rk} and offers the advantage that the
$\varepsilon$-scalar mass completely decouples from the physical
observables. For supersymmetric theories the \drbar{} and
\drbar{}$^\prime$ renormalization schemes are the same, while for
theories with broken  supersymmetry the latter one is most convenient.
At one-loop order there is only one relevant diagram
(cf. Fig.~\ref{fig::eps_prop}$(a)$) which has to be evaluated 
for vanishing external momentum.
A closer look to the two-loop diagrams shows that they
develop infra-red divergences in the limit $m_\varepsilon\to 0$
(cf., e.g., Fig.~\ref{fig::eps_prop}$(e)$). They can be regulated by 
introducing a small but non-vanishing mass for the 
$\varepsilon$ scalars. After the subsequent application
of an asymptotic expansion~\cite{Smirnov:2002pj}
in the limit $q^2=m_\varepsilon^2\ll M_q^2$ the infra-red divergences
manifest themselves as $\ln(m_\varepsilon)$ terms.
Furthermore, one-loop diagrams like the ones in 
Fig.~\ref{fig::eps_prop}$(b)$ and $(c)$ do not
vanish anymore and have to be taken into account as well.
Although they are proportional to $m_\varepsilon^2$, after
renormalization they induce two-loop contributions which
are proportional to $M_q^2$, partly multiplied by
$\ln(m_\varepsilon)$ terms. It is interesting to note that
in the sum of the genuine two-loop diagrams and the counterterm
contributions the limit $m_\varepsilon\to 0$ can be taken 
which demonstrates the infra-red finiteness of the on-shell mass of the 
$\varepsilon$ scalar.

Taking the infra-red finiteness for granted, it is also possible to 
choose $q^2=m_\varepsilon^2=0$ from the very beginning. 
Then the individual diagrams are 
infra-red divergent, however, the sum is not. 
We have performed the calculation both ways and checked that the final
result is the same. It is given by
\begin{eqnarray}
  \frac{M_q^2}{m_\varepsilon^2} Z_{m_\varepsilon}^{\rm OS} &=& 
  1 - \aepi n_h T_F
\bigg[
  \frac{2}{\ep} + 2 + 2 \Lmu + \ep\left(2 + \frac{1}{6} \pi^2  + 2 \Lmu
  + \Lmu^2 
\right)
\bigg] 
\nonumber\\
&& - \left(\apiDR\right)^2 n_h T_F
\bigg(\frac{3}{4} \frac{1}{\ep}  + \frac{1}{4} +
  \frac{3}{2} \Lmu
\bigg) C_A
 + \apiDR\aepi n_h T_F \bigg\{
\frac{1}{\ep^2} \left( \frac{3}{8} C_A + \frac{3}{2} C_F\right)
\nonumber\\
&& +\frac{1}{\ep}\bigg[ \frac{7}{8} C_A + \frac{3}{2} C_F +\left(
 \frac{3}{4} C_A  + \frac{3}{2} C_F\right)  \Lmu \bigg]
+ \left(\frac{15}{8} + \frac{1}{16}\pi^2 \right) C_A
\nonumber\\
&& + \left(\frac{3}{2} + \frac{1}{8}\pi^2\right) C_F + \left(\frac{7}{4}  C_A 
 +\frac{3}{2}  C_F\right)\Lmu +\left(\frac{3}{4} C_A + \frac{3}{4}
  C_F\right) \Lmu^2 
\bigg\}
\nonumber\\
&&
+ \left(\aepi\right)^2 n_h T_F \bigg\{
\frac{1}{\ep^2} \left(
\frac{1}{4} C_A  - \frac{1}{2} C_F  - \frac{1}{2} T_F n_f \right)
+\frac{1}{\ep}  \bigg[
\frac{1}{2} C_F
\nonumber\\
&& - \frac{1}{2} \left(1+\Lmu \right) T_F n_f  
\bigg]
- \frac{1}{2} C_A  + \frac{5}{2} C_F - \left(\frac{1}{2} +\frac{1}{24}
  \pi^2 \right) T_F n_f 
\nonumber\\
&&
- \left(\frac{1}{2} C_A - 2 C_F  + \frac{1}{2} T_F n_f \right) \Lmu 
- \left(\frac{1}{4} C_A  -\frac{1}{2} C_F  + \frac{1}{4} T_F n_f
  \right) \Lmu^2 
\bigg\}
\nonumber\\
&&
+ \aepi\frac{\eta_1}{\pi} n_h \bigg[
  \frac{3}{16} \frac{1}{\ep^2} +\frac{1}{\ep}\left(
\frac{3}{16} + \frac{3}{8} \Lmu \right) +\frac{3}{16}+\frac{1}{32} \pi^2 
+ \frac{3}{8} \Lmu + \frac{3}{8} \Lmu^2  
\bigg]
\nonumber\\
&&
- \aepi\frac{\eta_2}{\pi} n_h \bigg[
 \frac{5}{4} \frac{1}{\ep^2} + \left(5 + \frac{5}{2} \Lmu \right )\frac{1}{\ep}
+\frac{25}{2}  + \frac{5}{24} \pi^2 + 10 \Lmu + \frac{5}{2} \Lmu^2 
\bigg]
\nonumber\\
&&
- \aepi\frac{\eta_3}{\pi} n_h \bigg[
 \frac{7}{16}\frac{1}{\ep^2}  + \left( \frac{7}{16} + \frac{7}{8} \Lmu
  \right )\frac{1}{\ep}
+\frac{7}{16} + \frac{7}{96} \pi^2  + \frac{7}{8} \Lmu + \frac{7}{8}
  \Lmu^2 
\bigg]
  \,,
\end{eqnarray}
where the constants are defined after Eq.~(\ref{eq::mass::Zm2l}).
The overall factor $n_h$ in front of the one- and two-loop corrections
shows that the renormalization of $m_\varepsilon$ only influences 
those terms which contain a closed heavy quark loop.


\section{\label{sec::dimred}
  $Z_m^{\rm OS}$ in dimensional reduction}

The approach to extract the on-shell mass counterterm in \dred{} can be
taken over from \dreg{} as described in 
Section~\ref{sec::dimreg}, i.e. one considers the 
inverse quark propagator and requires that it has a zero at the
position of the pole. Again, the counterterm diagrams are generated
order-by-order in a generic way. 

\begin{figure}[t]
  \leavevmode
  \epsfxsize=\textwidth
  \epsffile{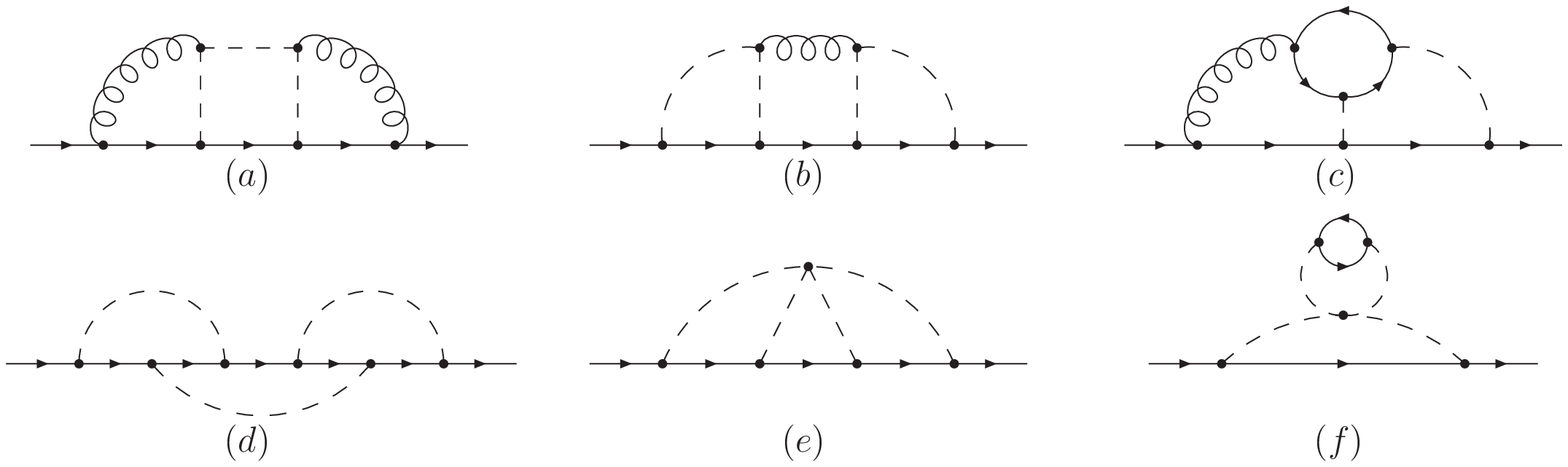}
\caption{\label{fig::selfdias_dred} Sample three-loop diagrams
  contributing to the quark propagator which have to be considered
  additionally in case dimensional reduction is used for the
  regularization. Solid lines
  denote massive quarks with mass $m_q$, curly lines denote gluons and 
  the $\varepsilon$ scalars are represented by dashed lines. In
  the closed fermion loops all quark flavours have to be considered.
  }
\end{figure}

The major complication as compared to the calculation in \dreg{} is
the appearance of the evanescent couplings and the $\varepsilon$
scalars. In particular, there are three different four-$\varepsilon$
vertices. This leads to many more Feynman diagrams which have
to be considered. Whereas in the case of \dreg{} about 130 diagrams
contribute there are more than 1100 in the case of \dred{}.
Typical Feynman diagrams are shown in Fig.~\ref{fig::selfdias}
and Fig.~\ref{fig::selfdias_dred}.
The more involved renormalization has already been discussed in
Section~\ref{sec::ren_dred}.

We avoid to present the result for the divergent quantity 
$Z_m^{\rm OS,\overline{\rm DR}}$ but show the result for the ratio
to the $\overline{\rm DR}$ quantity which is finite. We cast our
result in the following form
\begin{eqnarray}
  z_m^{\rm OS,\overline{\rm DR}} &=& 
  \frac{Z_m^{\rm OS,\overline{\rm DR}}}{Z_m^{\overline{\rm DR}}}
  \,\,=\,\, \frac{m_q^{\overline{\rm DR}}}{M_q^{\rm OS}} \,\,=\,\,
  1 
  + \delta^{(1)} z_m^{\rm OS,\overline{\rm DR}}
  + \delta^{(2)} z_m^{\rm OS,\overline{\rm DR}}
  + \delta^{(3)} z_m^{\rm OS,\overline{\rm DR}}
  \,.
  \label{eq::zmosdr}
\end{eqnarray}
Since our implementation of the couplings $\eta_i$ is only valid for
$SU(3)$ we furthermore set $N_c=3$. Our one-, two- and three-loop
results read:
\begin{eqnarray}
  \delta^{(1)} z_m^{\rm OS,\overline{\rm DR}} &=&
  - \apiDR \left( \frac{4}{3} + \Lmu \right)
  - \frac{1}{3} \aepi
  \,,
  \label{eq::zmdr1}
  \\
  \delta^{(2)} z_m^{\rm OS,\overline{\rm DR}} &=&
  \left( \apiDR \right)^2 \bigg[
    -\frac{3143}{288} - \frac{2}{9}\pi^2 - \frac{1}{9}\pi^2\ln2 +
    \frac{1}{6}\zeta_3 - \frac{151}{24}\Lmu - \frac{7}{8}\Lmu^2
  \nonumber\\
  && + \left( \frac{71}{144} + \frac{1}{18}\pi^2 + \frac{13}{36}\Lmu +
    \frac{1}{12}\Lmu^2 \right) n_l + \left( \frac{143}{144} -
    \frac{1}{9}\pi^2 + \frac{13}{36}\Lmu 
  \right.\nonumber\\&&\left.
  + \frac{1}{12}\Lmu^2 \right)
  n_h
  \bigg]
  + \apiDR\aepi \left( -\frac{3}{8} + \frac{1}{3}\Lmu \right) +
    \left( \aepi \right)^2 \left( \frac{1}{6} + \frac{1}{48}\, n_f
    \right)
  \,,
  \label{eq::zmdr2}
\end{eqnarray}
\begin{eqnarray}
  \delta^{(3)} z_m^{\rm OS,\overline{\rm DR}} &=&
  \left( \apiDR \right)^3
  \bigg\{
    -\frac{1160387}{10368} - \frac{24707}{2592}\pi^2 -
    \frac{38}{9}\pi^2\ln2 + \frac{7}{27}\pi^2\ln^22 +
    \frac{67}{72}\zeta_3
  \nonumber\\
  && + \frac{341}{2592}\pi^4 + \frac{1331}{432}\pi^2\zeta_3 -
    \frac{1705}{216}\zeta_5 + \frac{19}{54}\ln^42 + \frac{76}{9}a_4 -
    \bigg(
    \frac{20089}{288} + \pi^2
  \nonumber\\
  && + \frac{1}{2}\pi^2\ln2 - \frac{3}{4}\zeta_3
    \bigg)\Lmu
    - \frac{1475}{96}\Lmu^2 - \frac{21}{16}\Lmu^3 +
    \bigg[
      \frac{42235}{3888} + \frac{923}{648}\pi^2 + \frac{11}{81}\pi^2\ln2
  \nonumber\\
  && - \frac{2}{81}\pi^2\ln^22 + \frac{707}{216}\zeta_3 -
      \frac{61}{1944}\pi^4 - \frac{1}{81}\ln^42 - \frac{8}{27}a_4 +
      \bigg(
        \frac{3463}{432} + \frac{35}{108}\pi^2
  \nonumber\\
  && + \frac{1}{27}\pi^2\ln2 + \frac{7}{9}\zeta_3
      \bigg)\Lmu
      + \frac{35}{16}\Lmu^2 + \frac{2}{9}\Lmu^3
    \bigg] n_l -
    \bigg[
      \frac{2353}{23328} + \frac{13}{324}\pi^2 + \frac{7}{54}\zeta_3
  \nonumber\\
  && + \left( \frac{89}{648} + \frac{1}{54}\pi^2 \right)\Lmu +
      \frac{13}{216}\Lmu^2 + \frac{1}{108}\Lmu^3
    \bigg] n_l^2 -
    \bigg[
      \frac{5917}{11664} - \frac{13}{324}\pi^2 - \frac{2}{27}\zeta_3
  \nonumber\\
  && + \left( \frac{143}{324} - \frac{1}{54}\pi^2 \right)\Lmu +
    \frac{13}{108}\Lmu^2 + \frac{1}{54}\Lmu^3
    \bigg] n_ln_h +
    \bigg[
      \frac{77065}{3888} - \frac{13375}{1944}\pi^2
  \nonumber\\
  && + \frac{640}{81}\pi^2\ln2 + \frac{1}{81}\pi^2\ln^22 -
      \frac{751}{216}\zeta_3 - \frac{41}{972}\pi^4 +
      \frac{1}{4}\pi^2\zeta_3 - \frac{5}{4}\zeta_5 - \frac{1}{81}\ln^42
  \nonumber\\
  && - \frac{8}{27}a_4 + \left( \frac{4435}{432} - \frac{23}{54}\pi^2 +
      \frac{1}{27}\pi^2\ln2 + \frac{7}{9}\zeta_3 \right)\Lmu +
      \frac{35}{16}\Lmu^2 + \frac{2}{9}\Lmu^3
    \bigg] n_h
  \nonumber\\
  && -
    \bigg[
      \frac{9481}{23328} - \frac{4}{405}\pi^2 - \frac{11}{54}\zeta_3 +
      \left( \frac{197}{648} - \frac{1}{27}\pi^2 \right)\Lmu +
      \frac{13}{216}\Lmu^2 + \frac{1}{108}\Lmu^3
    \bigg] n_h^2
  \bigg\}
  \nonumber\\
  && + \left( \apiDR \right)^2 \aepi
  \bigg\{
    \frac{41105}{20736} + \frac{2}{27}\pi^2 + \frac{1}{27}\pi^2\ln2 +
    \frac{7}{24}\zeta_3 + \frac{35}{12}\Lmu + \frac{7}{24}\Lmu^2
  \nonumber\\
  && -
    \bigg[
      \frac{27}{64} + \frac{1}{54}\pi^2 + \frac{11}{54}\Lmu +
      \frac{1}{36}\Lmu^2 
    \bigg] n_l -
    \bigg[
      \frac{113}{192} - \frac{1}{27}\pi^2 + \frac{11}{54}\Lmu +
      \frac{1}{36}\Lmu^2 
    \bigg] n_h
  \bigg\}
  \nonumber\\
  && + \apiDR \left( \aepi \right)^2
  \bigg\{
    \frac{1397}{2592} - \frac{5}{36}\zeta_3 - \frac{1}{6}\Lmu +
    \bigg[
      \frac{55}{1728} + \frac{5}{36}\zeta_3 - \frac{1}{48}\Lmu
    \bigg] n_f
  \bigg\}
  \nonumber\\
  && + \left( \aepi \right)^3
  \bigg\{
    -\frac{7}{144} - \frac{5}{216}\zeta_3 - \frac{31}{576}\, n_f +
    \frac{5}{576}\, n_f^2
  \bigg\}
   - \frac{5}{24} \left( \aepi \right)^2 \frac{\eta_2}{\pi}
  \nonumber\\
  && - \frac{9}{256} \aepi \left( \frac{\eta_1}{\pi} \right)^2 +
  \frac{15}{16} \aepi \left( \frac{\eta_2}{\pi} \right)^2 -
  \frac{7}{128} \aepi \left( \frac{\eta_3}{\pi} \right)^2 +
  \frac{3}{64} \aepi \frac{\eta_1}{\pi} \frac{\eta_3}{\pi}
  \,.
  \label{eq::zmdr3}
\end{eqnarray}
For $\alpha_e=\alpha_s$ the two-loop result agrees
with~\cite{Avdeev:1997sz}, while the three-loop one is new.

There is a very strong cross check of the results
in Eqs.~(\ref{eq::zmdr1}),~(\ref{eq::zmdr2}) and~(\ref{eq::zmdr3})
for the limit $n_h=0$. The starting point is the relation 
between the \msbar{} and the on-shell mass in \dreg{} 
which can easily be
obtained from the results of Section~\ref{sec::dimreg} and the 
\msbar{} counterpart of Eq.~(\ref{eq::ZmDR}) (see, e.g.,
Ref.~\cite{Chetyrkin:2004mf}). 
In this relation, which depends on $\asMSbar$, both $\mMSbar$ and
$\asMSbar$ are replaced by their \dred{} counterparts using 
Eqs.~(4.2) and~(4.3) of Ref.~\cite{Harlander:2006xq}.
In this way we could verify the results for 
$z_m^{\overline{\rm DR},\rm OS}|_{n_h=0}$.
This provides a strong consistency check both on the results presented
in this paper but also on the approach used in~\cite{Harlander:2006xq}
for the extraction of the conversion formulae between the \msbar{} and
\drbar{} quantities.

In order to get an impression of the numerical size of the corrections,
both in \dreg{} and \dred{}, let us consider the relation between the
minimally subtracted and the pole mass for the case of the bottom and
top quark. As input we use $\alpha_s^{(5),\overline{\rm MS}}(M_Z) =
0.1189$~\cite{Bethke:2006ac}, $M_b=4.800$~GeV and
$M_t=171.4$~GeV~\cite{Group:2006xn}.

In the framework of \dreg{} it is straightforward to convert 
$\alpha_s^{(5),\overline{\rm MS}}(M_Z)$ to
$\alpha_s^{(5),\overline{\rm MS}}(M_b)$ and
$\alpha_s^{(6),\overline{\rm MS}}(M_t)$ using four-loop 
accuracy\footnote{We use {\tt RunDec}~\cite{Chetyrkin:2000yt} for the
  running and decoupling of $\alpha_s$ in the $\overline{\rm MS}$ scheme.}
leading to
\begin{align}
  & \mbox{bottom}: & m_b^{\msbarf}(M_b) = 3.953~\mbox{GeV} 
  =&\,\, 4.800 ( 1 - 0.0929 - 0.0493 - 0.0342)~\mbox{GeV} 
  \nonumber\\
  &&=&\,\, (4.800 - 0.445 - 0.236 - 0.164)~\mbox{GeV} 
  \,,
  \label{eq::mbmsexp}
  \\
  & \mbox{top}: & m_t^{\msbarf}(M_t) = 161.1~\mbox{GeV}  
  =&\,\, 171.4 ( 1 - 0.0461 - 0.0109 - 0.0033)~\mbox{GeV} 
  \nonumber\\
  &&=&\,\, (171.4 - 7.9 - 1.9 - 0.6 )~\mbox{GeV} 
  \,,
  \label{eq::mtmsexp}
\end{align}
with $\alpha_s^{(5),\overline{\rm MS}}(M_b)=0.2188$ and
$\alpha_s^{(6),\overline{\rm MS}}(M_t)=0.1085$.
The numbers given in the round brackets of the above equations
indicate the contributions from the tree-level, one-, two- and three-loop
conversion relation.

Within \dred{} the numerical analysis gets more involved since four
more couplings appear whose values are needed for $\mu=M_b$ and
$\mu=M_t$. As mentioned in the Introduction, \dred{} is an appropriate
scheme for supersymmetric theories where in the strong sector only
one coupling constant is present --- like in usual QCD using \dreg{}. 
Since all supersymmetric particles are heavier than the electroweak
scale it is necessary to match the full theory to the Standard Model
at some properly chosen scale $\mu^{\rm dec}$. At this step the
additional couplings appear.

For the computation of
$\alpha_e^{(5)}(M_b)$ and $\eta_i^{(5)}(M_b)$ we use $\mu^{\rm
  dec}=M_Z$, evaluate in a
first step $\alpha_s^{(5),\overline{\rm DR}}(M_Z)$ using the
three-loop relation given in Ref.~\cite{Harlander:2006xq} and require
\begin{eqnarray}
  &&\alpha_s^{(5),\overline{\rm DR}}(M_Z)
  = \alpha_e^{(5)}(M_Z) = \eta_1^{(5)}(M_Z)
  \,,\nonumber\\
  &&\eta_2^{(5)}(M_Z) = \eta_3^{(5)}(M_Z) = 0
  \,.
  \label{eq::as5ini}
\end{eqnarray}
In a second step the renormalization group functions, which are known
to four ($\alpha_s^{\overline{\rm DR}}$), three ($\alpha_e$)
and one-loop order ($\eta_i$)~\cite{Harlander:2006rj,Harlander:2006xq}
are used to obtain
\begin{eqnarray}
  \alpha_s^{(5),\overline{\rm DR}}(M_b) &=& 0.2241 \,,\nonumber\\
  \alpha_e^{(5)}(M_b) &=&  0.1721 \,,\nonumber\\
  \eta_1^{(5)}(M_b) &=&  0.2152 \,,\nonumber\\
  \eta_2^{(5)}(M_b) &=&  -0.01798 \,,\nonumber\\
  \eta_3^{(5)}(M_b) &=&  -0.005777 \,.
  \label{eq::asDRmb}
\end{eqnarray}

In the case of the top quark we choose $\mu^{\rm dec}=M_t$. It is
necessary to know the couplings for six active flavours and thus we
evaluate in a first step
$\alpha_s^{(6),\overline{\rm MS}}(M_Z)=0.1178$ and pose the analogue
requirements as in Eq.~(\ref{eq::as5ini}) with ``(5)'' replaced by
``(6)'' and $M_Z$ by $M_t$. This leads to
\begin{eqnarray}
  && \alpha_s^{(6),\overline{\rm DR}}(M_t) = 
  \alpha_e^{(6)}(M_t) = \eta_1^{(6)}(M_t) = 0.1096 \,,\nonumber\\
  && \eta_2^{(6)}(M_t) = \eta_3^{(6)}(M_t) = 0 \,.
  \label{eq::asDRmt}
\end{eqnarray}

Finally, we can insert the results of Eqs.~(\ref{eq::asDRmb})
and~(\ref{eq::asDRmt}) into Eq.~(\ref{eq::zmosdr}) and obtain
\begin{align}
  & \mbox{bottom}: & m_b^{\drbarf}(M_b) = 3.859~\mbox{GeV} 
  =&\,\, 4.800 ( 1 - 0.1134 - 0.0495 - 0.033)~\mbox{GeV}
  \nonumber\\
  &&=&\,\, (4.800 - 0.544 - 0.238 - 0.159)~\mbox{GeV} 
  \,, \\
  & \mbox{top}: & m_t^{\drbarf}(M_t) = 159.0~\mbox{GeV}  
  =&\,\, 171.4 ( 1- 0.0581  - 0.0105 - 0.0030)~\mbox{GeV} 
  \nonumber\\
  &&=&\,\, (171.4 - 10.0 - 1.8 - 0.5)~\mbox{GeV}  
  \,.
\end{align}
The perturbative expansion shows a similar behaviour as for the
$\overline{\rm MS}$--on-shell mass relation: the three-loop terms
amount to 159~MeV and 500~MeV, respectively, and are thus far above
the current uncertainty for the bottom quark mass~\cite{Kuehn:2007vp} and 
much larger than the expected uncertainty for the top quark
mass~\cite{Martinez:2002st}.


\section{Conclusions}

The main result of this paper is the relation between the pole and
\drbar{} quark mass to three-loop order in QCD where dimensional reduction
has been used as a regularization scheme.
The conversion formula has been obtained in analytical form where all
occurring integrals have been reduced to a small set of master
integrals with the help of the Laporta algorithm.
Due to the occurrence of evanescent couplings when using \dred{} within
QCD it is more advantageous to use in this case dimensional
regularization.
However, the latter cannot be used in supersymmetric models. Thus, 
our result constitutes an important preparation for similar
calculations in supersymmetric extensions of the Standard Model.

As a by-product we have obtained the corresponding relation and the
on-shell wave function renormalization using
dimensional regularization. This constitutes the first check of the 
analytical results obtained about seven years ago. Furthermore, we can
confirm the observation of Ref.~\cite{Melnikov:2000zc} that 
$Z_2^{\rm OS}$ depends on the gauge fixing parameter starting from
three-loop order.


\bigskip
\noindent
{\large\bf Acknowledgements}\\ 
We would like to thank Stefan Bekavac for discussions about the
Mellin-Barnes method and checking some of our results. 
This work was supported by the ``Impuls- und Vernetzungsfonds'' of the
Helmholtz Association, contract number VH-NG-008 and
the DFG through SFB/TR~9.
The Feynman diagrams were drawn with {\tt JaxoDraw} \cite{Binosi:2003yf}.




\begin{thebibliography}{99}

%
%

\bibitem{Bardeen:1978yd}
  W.~A.~Bardeen, A.~J.~Buras, D.~W.~Duke and T.~Muta,
  Phys.\ Rev.\  D {\bf 18} (1978) 3998.

\bibitem{Siegel:1979wr}
  W.~Siegel,
  Phys.\ Lett.\ B {\bf 84} (1979) 197.

\bibitem{Capper:1979ns}
  D.~M.~Capper, D.~R.~T.~Jones and P.~van Nieuwenhuizen,
  Nucl.\ Phys.\ B {\bf 167} (1980) 479.

\bibitem{Harlander:2006xq}
  R.~V.~Harlander, D.~R.~T.~Jones, P.~Kant, L.~Mihaila and M.~Steinhauser,
  JHEP {\bf 0612}, 024 (2006)
  [arXiv:hep-ph/0610206].

\bibitem{Tarrach:1980up}
  R.~Tarrach,
  Nucl.\ Phys.\  B {\bf 183} (1981) 384.

\bibitem{Gray:1990yh}
  N.~Gray, D.~J.~Broadhurst, W.~Grafe and K.~Schilcher,
  Z.\ Phys.\  C {\bf 48} (1990) 673.

\bibitem{Broadhurst:1991fy}
  D.~J.~Broadhurst, N.~Gray and K.~Schilcher,
  Z.\ Phys.\  C {\bf 52} (1991) 111.

\bibitem{Chetyrkin:1999ys}
  K.~G.~Chetyrkin and M.~Steinhauser,
  Phys.\ Rev.\ Lett.\  {\bf 83} (1999) 4001
  [arXiv:hep-ph/9907509].

\bibitem{Chetyrkin:1999qi}
  K.~G.~Chetyrkin and M.~Steinhauser,
  Nucl.\ Phys.\  B {\bf 573} (2000) 617
  [arXiv:hep-ph/9911434].

\bibitem{Melnikov:2000qh}
  K.~Melnikov and T.~van Ritbergen,
  Phys.\ Lett.\  B {\bf 482} (2000) 99
  [arXiv:hep-ph/9912391].

\bibitem{Melnikov:2000zc}
  K.~Melnikov and T.~van Ritbergen,
  Nucl.\ Phys.\ B {\bf 591} (2000) 515
  [arXiv:hep-ph/0005131].

\bibitem{Martin:1993yx}
  S.~P.~Martin and M.~T.~Vaughn,
  Phys.\ Lett.\ B {\bf 318} (1993) 331
  [arXiv:hep-ph/9308222].

\bibitem{Avdeev:1997sz}
  L.~V.~Avdeev and M.~Y.~Kalmykov,
  Nucl.\ Phys.\ B {\bf 502} (1997) 419
  [arXiv:hep-ph/9701308].

\bibitem{Marquard:2006qi}
  P.~Marquard, J.~H.~Piclum, D.~Seidel and M.~Steinhauser,
  Nucl.\ Phys.\ B {\bf 758} (2006) 144
  [arXiv:hep-ph/0607168].

\bibitem{Piclum_diss}
  J.~H.~Piclum,	Dissertation, Universit\"at Hamburg,
      in preparation.

\bibitem{Nogueira:1991ex}
  P.~Nogueira,
  J.\ Comput.\ Phys.\  {\bf 105} (1993) 279.

\bibitem{Harlander:1997zb}
  R.~Harlander, T.~Seidensticker and M.~Steinhauser,
  Phys.\ Lett.\ B {\bf 426} (1998) 125
  [hep-ph/9712228].

\bibitem{Seidensticker:1999bb}
  T.~Seidensticker,
  hep-ph/9905298.

\bibitem{Laporta:1996mq}
  S.~Laporta and E.~Remiddi,
  Phys.\ Lett.\ B {\bf 379} (1996) 283
  [arXiv:hep-ph/9602417].

\bibitem{Laporta:2001dd}
  S.~Laporta,
  Int.\ J.\ Mod.\ Phys.\ A {\bf 15} (2000) 5087
  [arXiv:hep-ph/0102033].

\bibitem{PMDS}
  P.~Marquard and D.~Seidel,
  unpublished.

\bibitem{Bauer:2000cp}
  C.~Bauer, A.~Frink and R.~Kreckel,
  arXiv:cs.sc/0004015.

\bibitem{fermat} R.~H.~Lewis, Fermat's User Guide,
  http://www.bway.net/\~{}lewis.

\bibitem{Tentyukov:2006ys}
  M.~Tentyukov and J.~A.~M.~Vermaseren,
  arXiv:cs.sc/0604052.

\bibitem{Chetyrkin:1981qh}
  K.~G.~Chetyrkin and F.~V.~Tkachov,
  Nucl.\ Phys.\ B {\bf 192} (1981) 159.
%

\bibitem{Smirnov:1999gc}
  V.~A.~Smirnov,
  Phys.\ Lett.\ B {\bf 460} (1999) 397
  [arXiv:hep-ph/9905323].

\bibitem{Tausk:1999vh}
  J.~B.~Tausk,
  Phys.\ Lett.\ B {\bf 469} (1999) 225
  [arXiv:hep-ph/9909506].

\bibitem{Czakon:2005rk}
  M.~Czakon,
  Comput.\ Phys.\ Commun.\  {\bf 175} (2006) 559
  [arXiv:hep-ph/0511200].

\bibitem{Jack:1993ws}
  I.~Jack, D.~R.~T.~Jones and K.~L.~Roberts,
  Z.\ Phys.\ C {\bf 62} (1994) 161
  [arXiv:hep-ph/9310301].

\bibitem{Harlander:2006rj}
  R.~Harlander, P.~Kant, L.~Mihaila and M.~Steinhauser,
  JHEP {\bf 0609}, 053 (2006)
  [arXiv:hep-ph/0607240].

\bibitem{Jack:1994rk}
  I.~Jack, D.~R.~T.~Jones, S.~P.~Martin, M.~T.~Vaughn and Y.~Yamada,
  Phys.\ Rev.\ D {\bf 50} (1994) 5481
  [arXiv:hep-ph/9407291].

\bibitem{Smirnov:2002pj}
  V.~A.~Smirnov,
  ``Applied asymptotic expansions in momenta and masses,''
  Springer Tracts Mod.\ Phys.\  {\bf 177} (2002) 1.

\bibitem{Chetyrkin:2004mf}
  K.~G.~Chetyrkin,
  Nucl.\ Phys.\  B {\bf 710} (2005) 499
  [arXiv:hep-ph/0405193].

\bibitem{Bethke:2006ac}
  S.~Bethke,
  arXiv:hep-ex/0606035.

\bibitem{Group:2006xn}
  Tevatron Electroweak Working Group,
  hep-ex/0608032.

\bibitem{Chetyrkin:2000yt}
  K.~G.~Chetyrkin, J.~H.~K\"uhn and M.~Steinhauser,
  Comput.\ Phys.\ Commun.\  {\bf 133} (2000) 43
  [arXiv:hep-ph/0004189].

\bibitem{Kuehn:2007vp}
  J.~H.~K\"uhn, M.~Steinhauser and C.~Sturm,
  arXiv:hep-ph/0702103.

\bibitem{Martinez:2002st}
  M.~Martinez and R.~Miquel,
  Eur.\ Phys.\ J.\  C {\bf 27} (2003) 49
  [arXiv:hep-ph/0207315].

\bibitem{Binosi:2003yf}
  D.~Binosi and L.~Theussl,
  Comput.\ Phys.\ Commun.\  {\bf 161} (2004) 76
  [arXiv:hep-ph/0309015].


\end{thebibliography}
\end{document}